\documentclass{aa} 
\usepackage{graphicx}
\usepackage{multirow}
\usepackage{subcaption}

\begin{document}
\title{Magnetohydrodynamic waves in braided magnetic fields}
\author{T. A. Howson \inst{1} \and I. De Moortel \inst{1,2} \and J. Reid \inst{1} \and A. W. Hood \inst{1}}
\institute{School of Mathematics and Statistics, University of St Andrews, St Andrews, Fife, KY16 9SS, U.K. \and Rosseland Centre for Solar Physics, University of Oslo, PO Box 1029  Blindern, NO-0315 Oslo, Norway}

\abstract{}
{We investigate the propagation of transverse magnetohydrodynamic (MHD) wave fronts through a coronal plasma containing a braided magnetic field.}
{We performed a series of three dimensional MHD simulations in which a small amplitude, transverse velocity perturbation is introduced into a complex magnetic field. We analysed the deformation of the wave fronts as the perturbation propagates through the braided magnetic structures and explore the nature of Alfv\'enic wave phase mixing in this regime. We considered the effects of viscous dissipation in a weakly non-ideal plasma and evaluate the effects of field complexity on wave energy dissipation.}  
{Spatial gradients in the local Alfv\'en speed and variations in the length of magnetic field lines ensure that small scales form throughout the propagating wave front due to phase mixing. Additionally, the presence of complex, intricate current sheets associated with the background field locally modifies the polarisation of the wave front. The combination of these two effects enhances the rate of viscous dissipation, particularly in more complex field configurations. Unlike in classical phase mixing configurations, the greater spatial extent of Alfv\'en speed gradients ensures that wave energy is deposited over a larger cross-section of the magnetic structure. Further, the complexity of the background magnetic field ensures that small gradients in a wave driver can map to large gradients within the coronal plasma.}    
{The phase mixing of transverse MHD waves in a complex magnetic field will progress throughout the braided volume. As a result, in a non-ideal regime wave energy will be dissipated over a greater cross-section than in classical phase mixing models. The formation rate of small spatial scales in a propagating wave front is a function of the complexity of the background magnetic field. As such, if the coronal field is sufficiently complex it remains plausible that phase mixing induced wave heating can contribute to maintaining the observed temperatures. Furthermore, the weak compressibility of the transverse wave and the observed phase mixing pattern may provide seismological information about the nature of the background plasma.}
{}
\maketitle

%%%%%%%%%%%%%%%%%%%%%%%%%%%%%%%%%%%%%%%%%%%%%%%%%%%%%%%%%%%%%%%%%%%%%%%
%Introduction
%%%%%%%%%%%%%%%%%%%%%%%%%%%%%%%%%%%%%%%%%%%%%%%%%%%%%%%%%%%%%%%%%%%%%%%
\section{Introduction}\label{sec:introduction}
Over recent years, high spatial and temporal resolution observations have allowed many authors to identify the existence of a wide variety of magnetohydrodynamic (MHD) waves throughout the solar corona \citep[e.g.][]{Aschwanden1999, Aschwanden2002, Verwichte2010, Threlfall2013, Duckenfield2018}. Of particular interest to the current study is the evidence of waves propagating along coronal structures \cite[e.g.][]{DeMoortel2000, McEwan2006, Okamoto2007, Thurgood2014, Morton2015}. Whilst estimates of the energy associated with these waves are not well constrained \citep[see, for example,][]{Tomczyk2007, McIntosh2011}, it is hypothesised that they may contribute to the heating of the coronal plasma and/or the acceleration of the fast solar wind. We refer interested readers to reviews by \citet{Erdelyi2007,DeMoortel2012, Arregui2015}, and references therein.

Whilst the shuffling of magnetic foot points by photospheric motions may be the source of propagating MHD waves \citep[e.g.][]{Cranmer2005, Engvold2008, Wang2009, Hillier2013}, flows at the solar surface may also cause the slow braiding of the coronal magnetic field \citep{Parker1972}. This may lead to the formation of coronal current sheets and the heating of plasma through Ohmic dissipation and magnetic reconnection, even in a weakly non-ideal regime \citep[e.g.][]{Hardi2004, Klimchuk2006, WilmotSmith2011, Reale2016, OHara2016}. It is hypothesised that the coronal magnetic field may exist in a permanently stressed state and, as such, the background structures that are perturbed by MHD waves may have a complex and highly structured topology \citep[e.g.][]{Longcope1994, Longbottom1998, WilmotSmith2015, Pontin2016}. Indeed, whilst the small scale nature of the coronal field is not well-constrained, many observational studies have highlighted the presence of large, stressed magnetic structures and posited their association with impulsive energy release \citep[e.g.][]{Srivastava2010, Yan2014, Joshi2015, Lim2016}. 

Many previous authors have modelled the propagation of waves through a variety of coronal-like plasmas containing a range of magnetic field configurations. These include considerations of wave behaviour around magnetic null points \citep[e.g.][]{McLaughlin2016, McLaughlin2019, Prok2019}, wave interactions with large scale fields \citep[e.g.][]{Ofman2002, Afanasyev2018}, and oscillations in randomly structured plasmas \citep[e.g.][]{Pascoe2011, Yuan2015, Yuan2016, Magyar2017}. Unlike in uniform media, the fast and Alfv\'en modes are not decoupled in inhomogeneous plasmas and waves with mixed properties are able to propagate \citep[e.g.][]{Goossens2009, Goossens2013}. For example, in cylindrical geometries, there is an extensive body of literature investigating the kink mode and its resonant coupling to azimuthal Alfv\'en waves \citep[][]{Ionson1978, Goossens2011}. In a more general setting, \citet{Lazzaro2000}, present a method for investigating the propagation of MHD waves through finely structured plasmas.

The propagation of waves through a plasma with a transverse gradient in the Alfv\'en speed can lead to phase mixing \citep{Heyvaerts1983} as wave fronts on adjacent magnetic field lines travel at different velocities. As such, large spatial gradients in the perturbed magnetic and velocity fields can form and will enhance the rate of wave heating dissipation even in the weakly dissipative coronal plasma \citep[e.g.][]{Pagano2017}. However, in \citet{Cargill2016}, the authors highlighted concerns of the suitability of the classical phase mixing regime for heating the corona. In particular, they argued that the typically assumed density profile cannot be self-consistently sustained without an additional background heating term. Additionally, they show that the rate of wave energy dissipation is insufficient unless transport coefficients well in excess of expected coronal values are implemented. 

In the classical phase mixing regime, the dissipation time scale is expected to be proportional to $R_t^{^{1}/_{3}}$, where $R_t$ is a combination of the fluid and magnetic Reynolds numbers \citep{Heyvaerts1983}. However, an analytic investigation of Alfv\'en waves in more complex magnetic field structures presented in \citet{Similon1989}, demonstrates that in some geometries, the dissipation time may be proportional to $\log{R_M}$. This may provide a significant enhancement above the classical heating rate and suggests phase mixing may be an important wave energy dissipation mechanism if the coronal field is sufficiently complex.   

In this article, we present the results of numerical models investigating the propagation of MHD waves through a braided magnetic field. We investigate the effects of the field complexity, viscous and Ohmic dissipation, and numerical resolution. In Section 2, we outline our numerical model and describe the initial conditions of the simulations. In Section 3 we present our results and finally, in Section 4, we present a discussion of the implications of this study.

%%%%%%%%%%%%%%%%%%%%%%%%%%%%%%%%%%%%%%%%%%%%%%%%%%%%%%%%%%%%%%%%%%%%%%%
%Numerical Method
%%%%%%%%%%%%%%%%%%%%%%%%%%%%%%%%%%%%%%%%%%%%%%%%%%%%%%%%%%%%%%%%%%%%%%%
\section{Numerical method}
\subsection{Initial conditions}
For the numerical simulations presented within this article, we seek initial conditions consisting of complex magnetic field topologies. In order to obtain such fields, we have used the output of a 3D MHD model presented by \citet{Reid2018}. Whilst the results of the simulations presented in the aforementioned paper are not directly related to the current work, we present a brief description of the model in order to describe the initial conditions of our experiments.

\begin{figure}[h]
  \centering
  \includegraphics[width=0.5\textwidth]{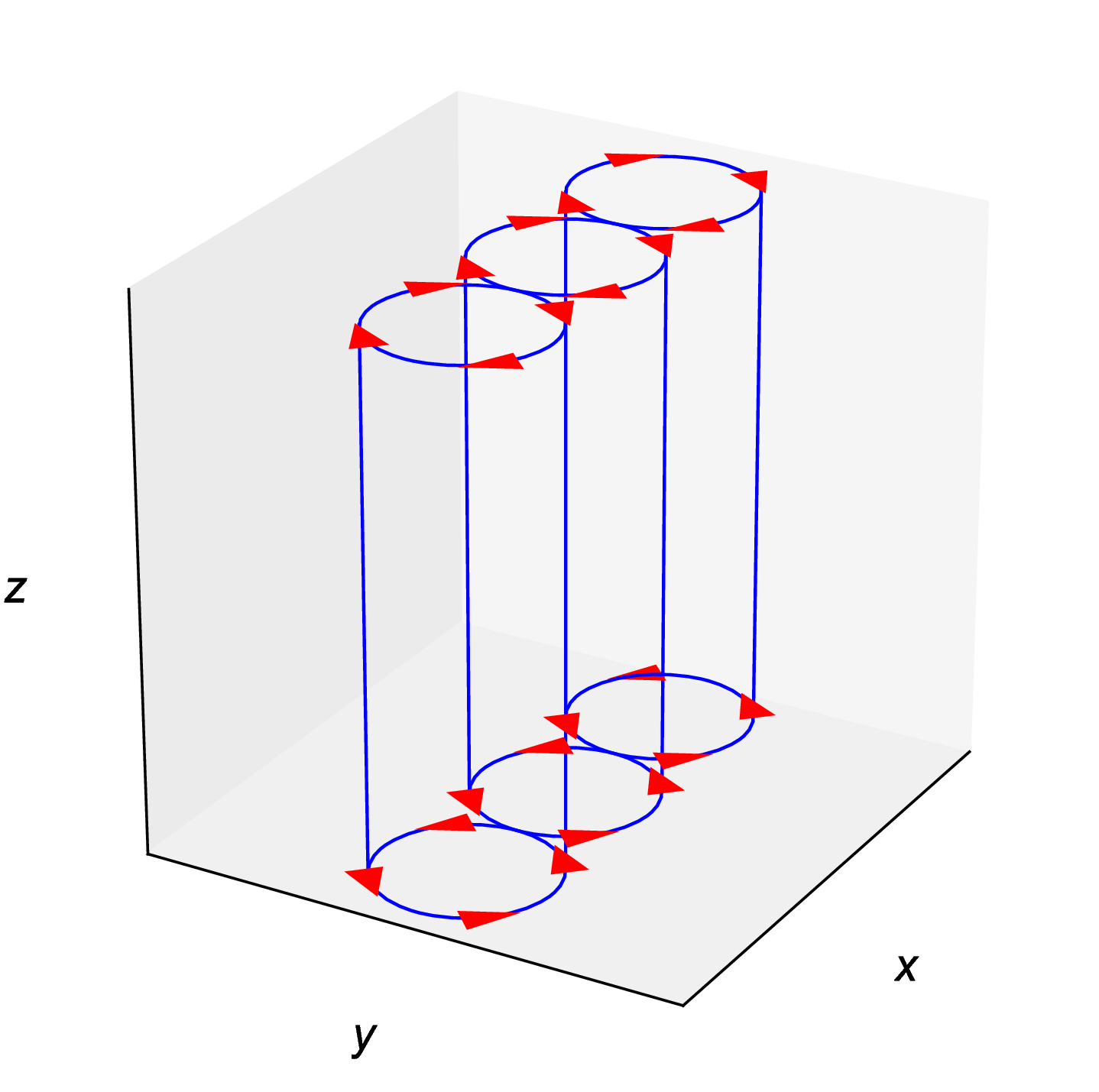}
  \caption{Schematic of rotational driving implemented in \citet{Reid2018}.}
  \label{Drive_cartoon}
\end{figure}

In \citet{Reid2018}, an initially uniform, three dimensional, straight magnetic field is stressed by continuous, counter-rotational motions imposed on both foot points of three cylindrical threads. The configuration of the velocity driver is shown in Fig. \ref{Drive_cartoon}. These motions generate columns of twisted field, which, at least initially, remain distinct. The central thread is twisted at a faster rate than the other two and hence, reaches the threshold for kink instability sooner. The onset of the instability generates complex current structures and destabilises the remaining two magnetic threads. 

Ultimately, the constant injection of Poynting flux by the continuous rotational driving generates distributed current sheets and a complex field structure throughout the domain. In a non-ideal regime, viscous and Ohmic heating increases the plasma temperature during the course of the simulation. We direct the interested reader to the analysis presented in \citet{Reid2018} for further details.

In the present study, we used four simulation times from the model described above in order to obtain fields with varying field complexity and different quantities of magnetic twist. In each case we relaxed the field structure towards a numerical equilibrium under the action of a large viscosity. This causes further plasma heating but allows the wave dynamics (see Results) to be analysed separately from the velocities associated with the previous rotational driving. The numerical relaxation continues until the velocities are small in comparison to the amplitude of the wave driver (see below).

As in \citet{Reid2018}, for the numerical simulations presented within this paper, we have implemented the Lagrangian-remap code, Lare3D \citep{Arber2001}. We advance the full, 3D  MHD equations in normalised form given by

\begin{equation}\frac{\text{D}\rho}{\text{D}t} = -\rho \vec{\nabla} \cdot \vec{v}, \end{equation}
\begin{equation} \label{eq:motion} \rho \frac{{\text{D}\vec{v}}}{{\text{D}t}} = \vec{j} \times \vec{B} - \vec{\nabla} P + \vec{F}_{\text{visc.}}, \end{equation}
\begin{equation} \label{eq:energy} \rho \frac{{\text{D}\epsilon}}{{\text{D}t}} = \eta \lvert \vec{j}\rvert^2- P(\vec{\nabla} \cdot \vec{v}) + Q_{\text{visc.}}, \end{equation}
\begin{equation}\frac{\text{D}\vec{B}}{\text{D}t}=\left(\vec{B} \cdot \vec{\nabla}\right)\vec{v} - \left(\vec{\nabla} \cdot \vec{v} \right) \vec{B} - \vec{\nabla} \times \left(\eta \vec{\nabla} \times \vec{B}\right), \end{equation}
where all variables have their usual meanings. The term $\vec{F}_{\text{visc.}}$ is the combination of viscous forces and $Q_{\text{visc.}}$ is the associated heating. These dissipative terms can be separated into contributions from a \emph{real} viscosity, $\nu$, and two small shock viscosity terms which are included within the following simulations to ensure numerical stability. Unless otherwise stated $\nu$ is set to zero (see section \ref{Diss_sec} for discussion of non-zero viscosity).

Our numerical domain has dimensions of 30 Mm $\times$ 30 Mm $\times$ 100 Mm (in $x$, $y$ and $z$, respectively) and in the majority of the simulations presented below, we implemented a numerical grid of 128 $\times$ 128 $\times$ 512. We also considered a higher resolution case with 256 $\times$ 256 $\times$ 1024 grid cells. The equilibrium obtained during the relaxation phase is resolution dependent and narrower current sheets are present when the more refined grid is used (see below for more details).

The model presented in \citet{Reid2018} is driven continuously and the field complexity slowly increases during the course of the simulations. In order to attain various degrees of field complexity, we use the simulation states at 100, 200, 300 and 500 Alfv\'en times as the initial conditions for the numerical relaxation. Hereafter, we refer to these simulations as s1, s2, s3 and s5 (t5 for the high resolution simulation), respectively. In each case, the rotational driving is ceased in order to allow the field to relax.

\begin{figure}[h]
  \centering
  \includegraphics[width=0.5\textwidth]{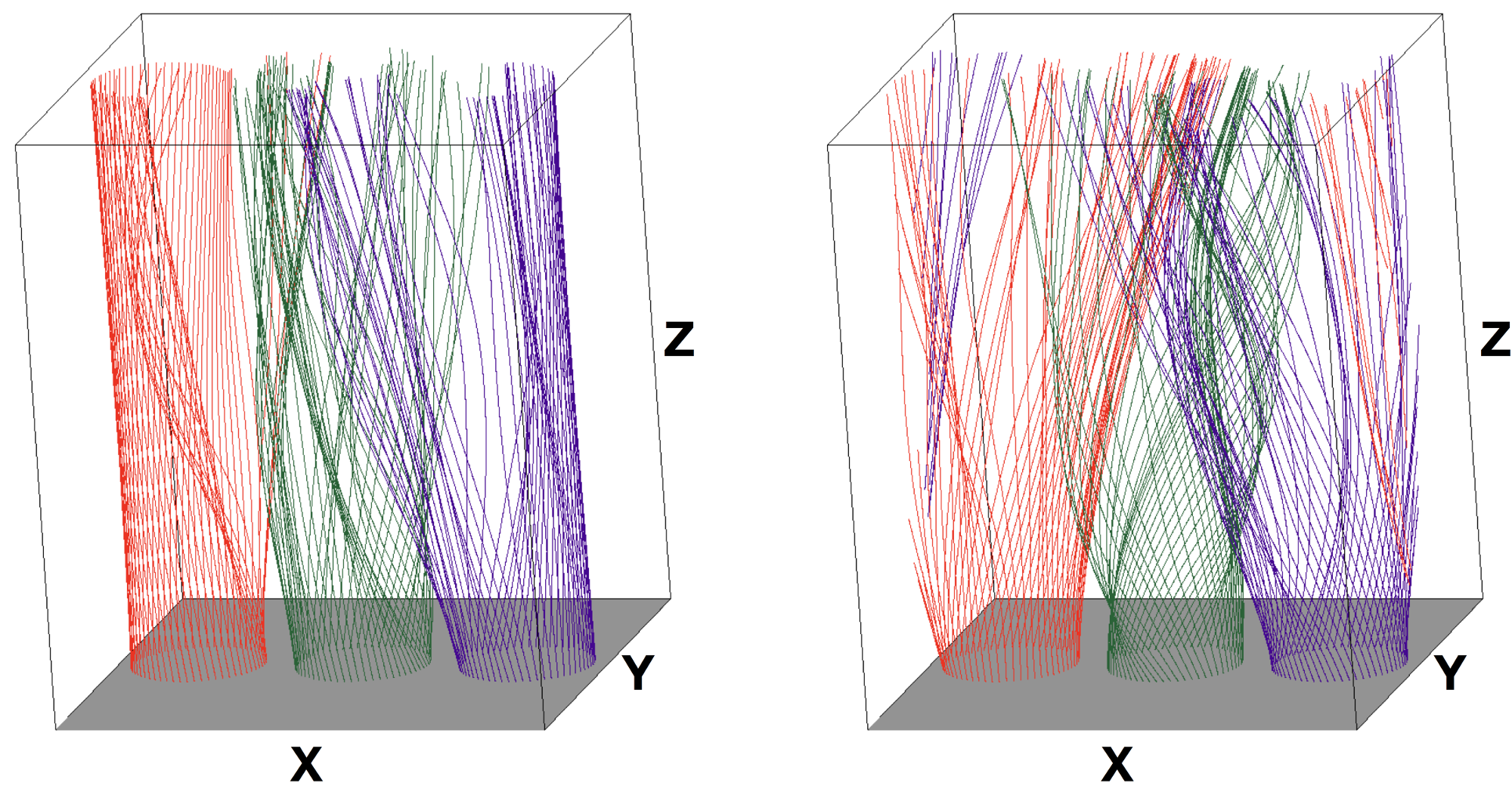}
  \caption{Configuration of magnetic field lines for the least (s1; left-hand panel) and most (s5; right-hand panel) complex initial conditions.}
  \label{In_Field_Vertical}
\end{figure}

\begin{figure}[h]
  \centering
  \includegraphics[width=0.5\textwidth]{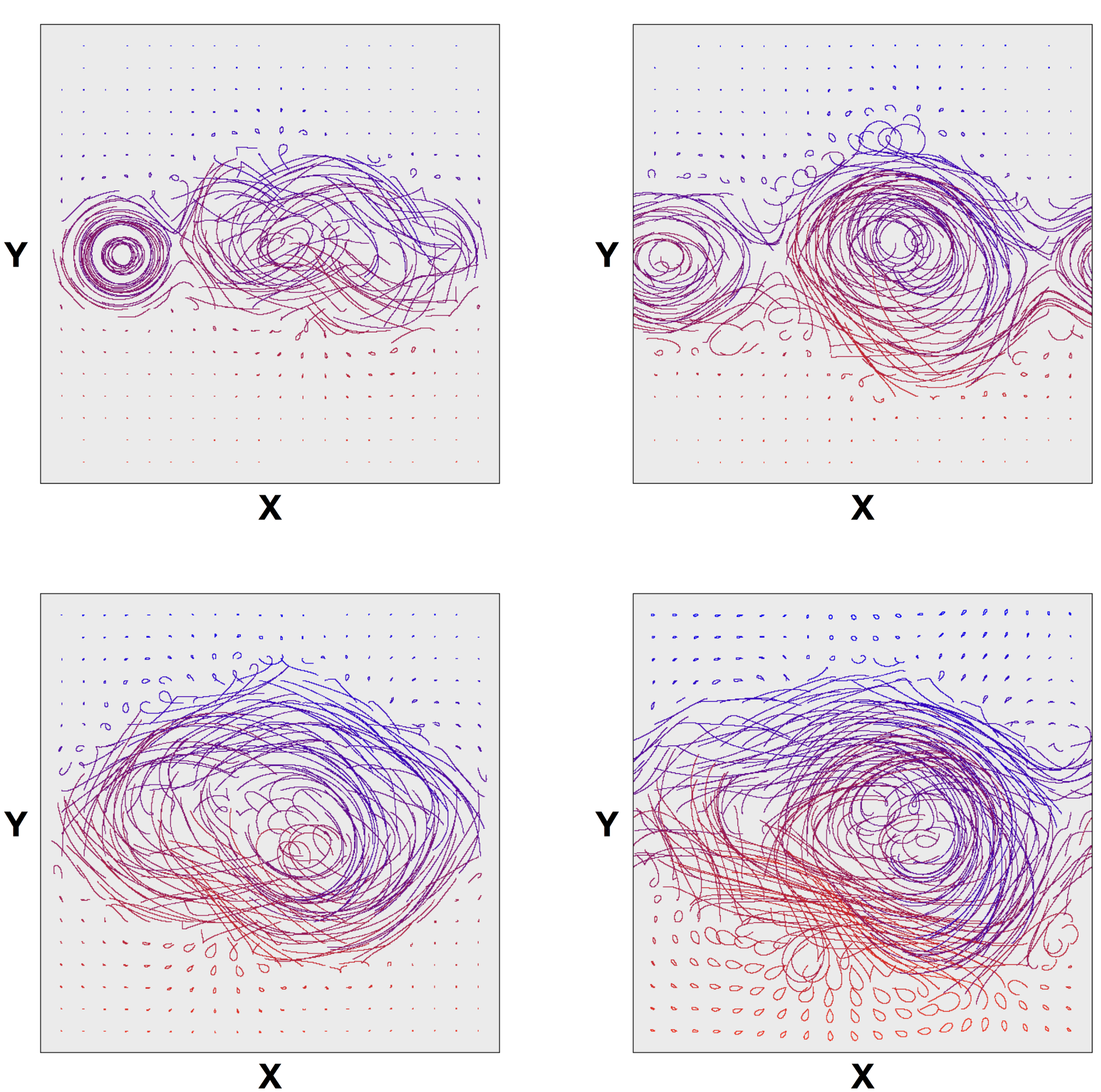}
  \caption{Projections of magnetic field lines onto $x$-$y$-plane for the four field configurations. {\emph{Upper left:}} s1, {\emph{upper right:}} s2, {\emph{lower left:}} s3 and {\emph{lower right:}} s5.}
  \label{In_Field_Top}
\end{figure}

In Figs. \ref{In_Field_Vertical} and \ref{In_Field_Top}, we show the configuration of the magnetic field for various cases following the numerical relaxation. In Fig. \ref{In_Field_Vertical}, magnetic field lines are traced from the foot points of each cylinder for the least (s1; left-hand panel) and most (s5; right-hand panel) twisted simulations. The colour of each field line identifies the cylinder containing the lower foot point of the field line. In Fig. \ref{In_Field_Top}, we display the projections of magnetic field lines onto the lower boundary of the domain for each of the four post-relaxation states. In this case, the colour of the field line is simply for visualisation purposes. In the least braided simulation (upper left panel), we see that the central and right-hand flux tube have both become unstable and have merged to form a larger structure. However, the left-hand flux tube has remained distinct and we simply observe the helical nature of the constituent field lines. For the field configurations that have been stressed further (see lower row of Fig. \ref{In_Field_Top} and right-hand panel of Fig. \ref{In_Field_Vertical}), the three cylinders are all inter-connected and no separate structures remain. 

\begin{figure}
\centering
\begin{subfigure}[b]{0.2\textwidth}
   \includegraphics[width=1\linewidth]{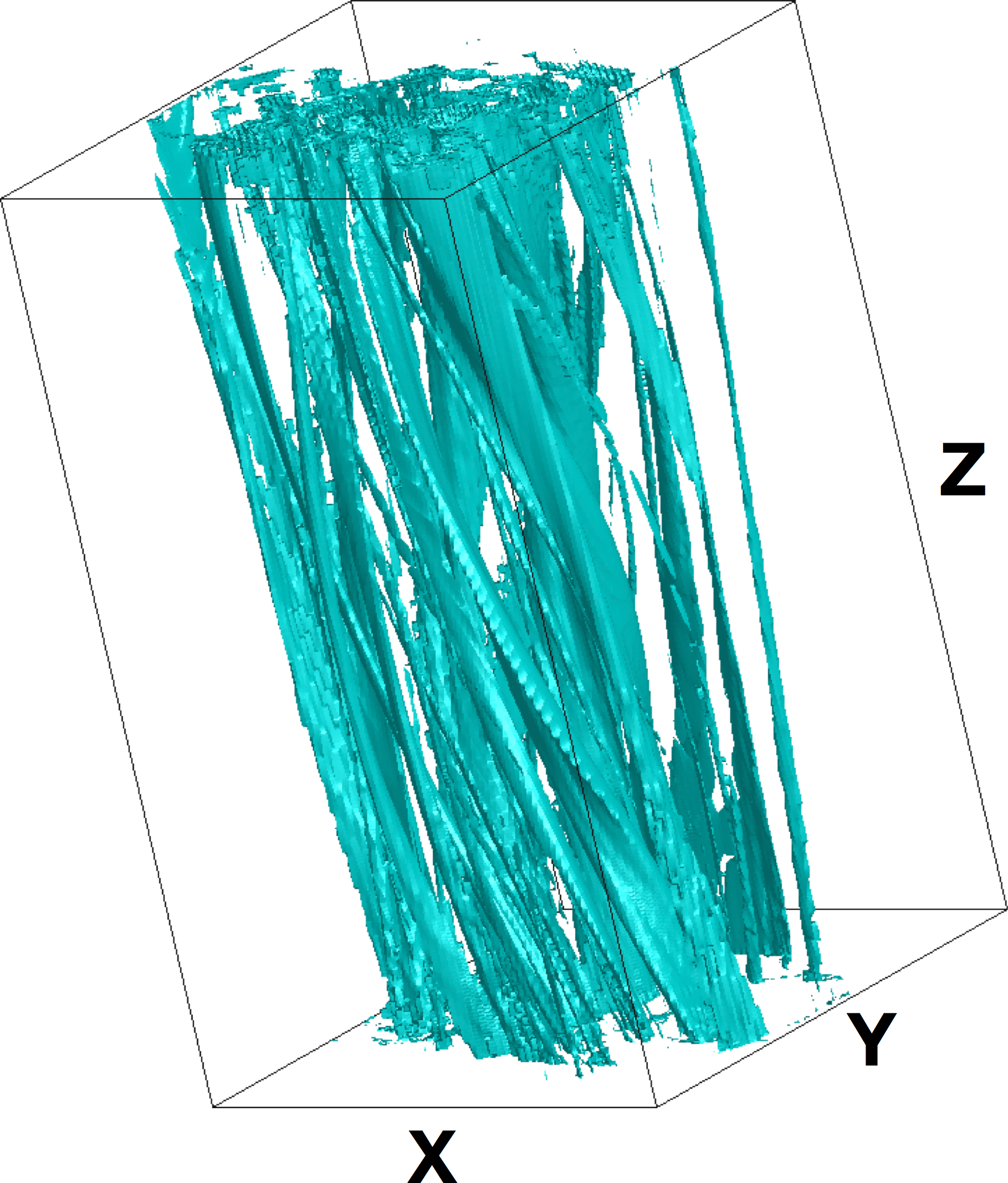}
   \caption{Current.}
   \label{In_cur}
\end{subfigure}
\begin{subfigure}[b]{0.2\textwidth}
   \includegraphics[width=1\linewidth]{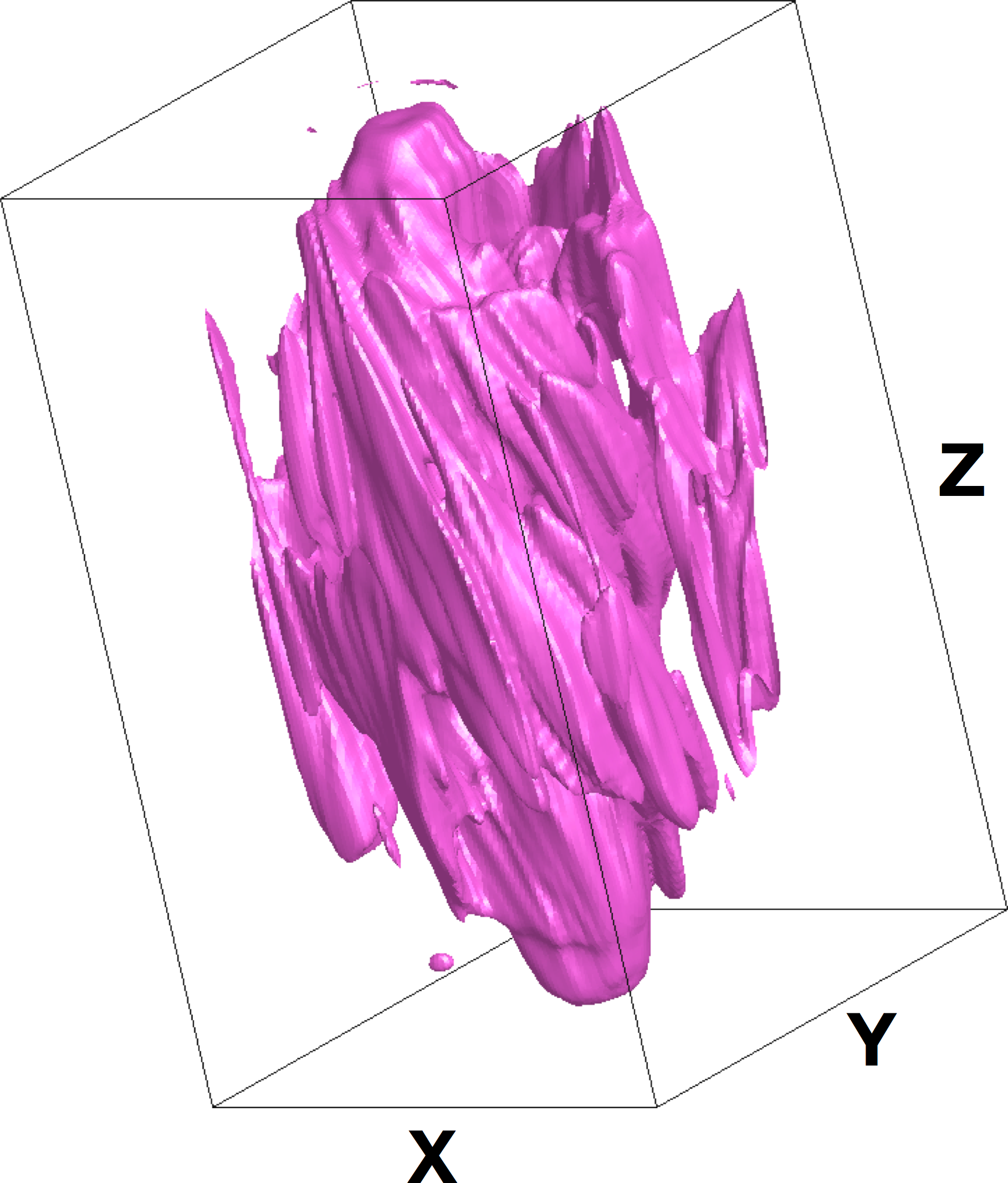}
   \caption{Alfv\'en Speed.}
   \label{In_alf} 
\end{subfigure}
\caption{Initial configuration for the most complex field (s5) simulation. In the right-hand panel, the Alfv\'en speed is largest within the volume.}
  \label{In_structure}
\end{figure}

In all cases, the resultant plasma is highly inhomogeneous with spatial gradients in the Alfv\'en speed, gas and magnetic pressures, density and temperature. The mean, minimum and maximum values of plasma parameters in the relaxed states are presented in Table \ref{Tab_eqm_values}. Due to the plasma heating described in \citet{Reid2018}, and because of the viscous heating that occurs during the relaxation phase, the more complex field simulations exhibit an increase in the mean temperature. We note that the very high temperatures observed in the more complex field simulations (s2-s5 and t5) are typically confined to a few grid points. These locations have been heated by the most energetic current sheets and are not representative of the conditions throughout the remainder of the domain. Despite the inhomogeneity present within much of the numerical grid, for large values of $|y|$, the field remains relatively simple (see Fig. \ref{In_Field_Top}) as the plasma was not exposed to rotational driving in these locations (see Fig. \ref{Drive_cartoon}).

In each case $\beta < 1$, and the relaxed field is approximately force-free. Since $\vec{j} \times \vec{B} \approx \vec{0}$, the currents are dominated by the component parallel to $\vec{B}$. As the relaxed, approximately force-free field is predominantly aligned with the $z$ axis, we have $j_x \sim j_y \ll j_z$. This property of the background currents is important for explaining the modification in wave polarisation discussed in Section \ref{Results_section1}. An isosurface of the initial current associated with the most complex field (s5) is displayed in the left-hand panel of Fig. \ref{In_structure}.  

In the right-hand panel of Fig. \ref{In_structure}, we display an isosurface of the Alfv\'en speed for the same simulation. The viscous and Ohmic heating generated by the rotational driving is largest close to the centre of the domain. As such, the temperature increases most rapidly here. Therefore, in order to maintain pressure balance, density is expelled (predominantly along magnetic field lines) from the centre of the domain and thus the Alfv\'en speed is higher here. 

\begin{figure}[h]
  \centering
  \includegraphics[width=0.5\textwidth]{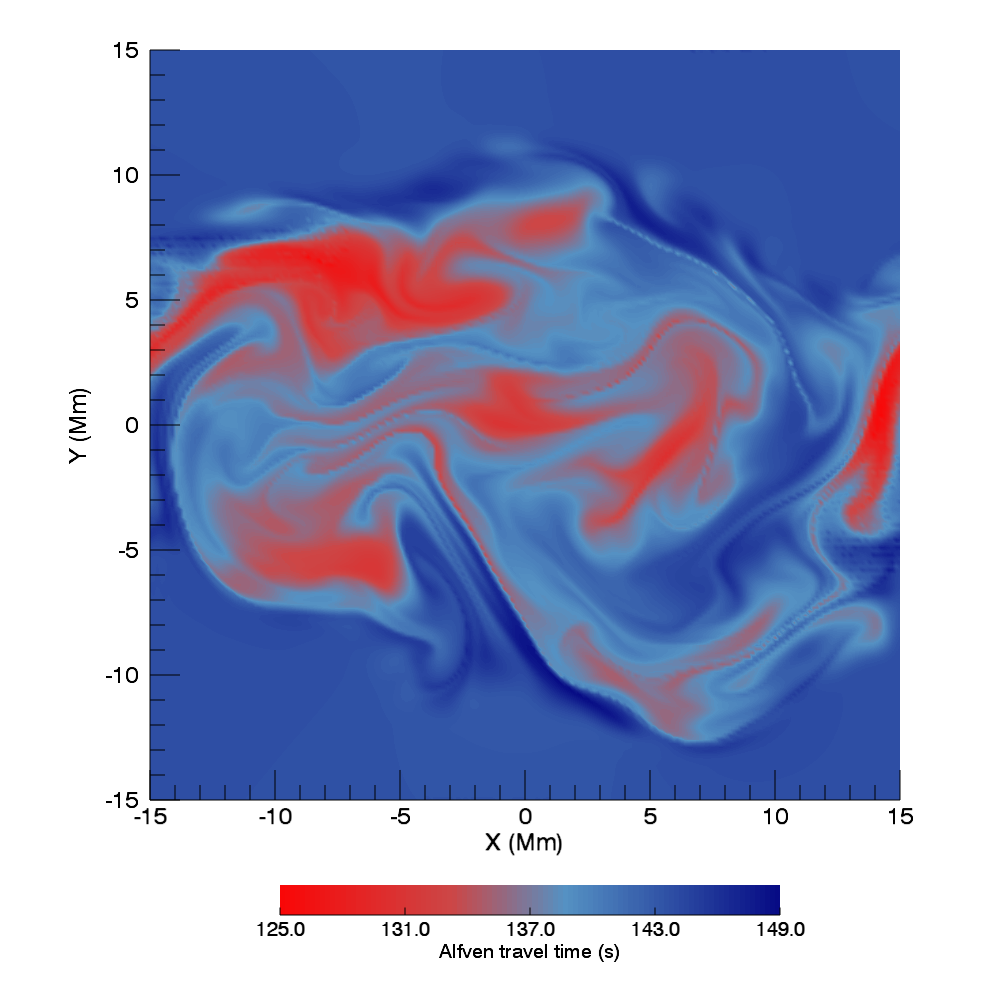}
  \caption{The Alfv\'en travel time, $\Omega$ (see equation \ref{eq_Alf_travel}), along magnetic field lines traced from the upper $z$ boundary.}
  \label{alf_travel_time}
\end{figure}

Despite this, the Alfv\'en travel time is not necessarily shorter in the centre of the $x$-$y$-plane, as here, field lines are typically more twisted, and hence, longer. In Fig. \ref{alf_travel_time}, we display the travel time for a wave propagating at the local Alfv\'en speed along magnetic field lines traced from the upper $z$ boundary. In particular we calculate 
\begin{equation} \label{eq_Alf_travel} \Omega(x,y) = \int_{z_{\text{min}}}^{z_{\text{max}}} \frac{\mathrm{d} s}{v_A}, \end{equation} 
where $s$ is an infinitesimal field line element. The spatial scales in the travel time, $\Omega$, provide an estimate for the amount of phase mixing that can be expected as a wave front propagates through the complex field structures. 

\begin{figure}[h]
  \centering
  \includegraphics[width=0.5\textwidth]{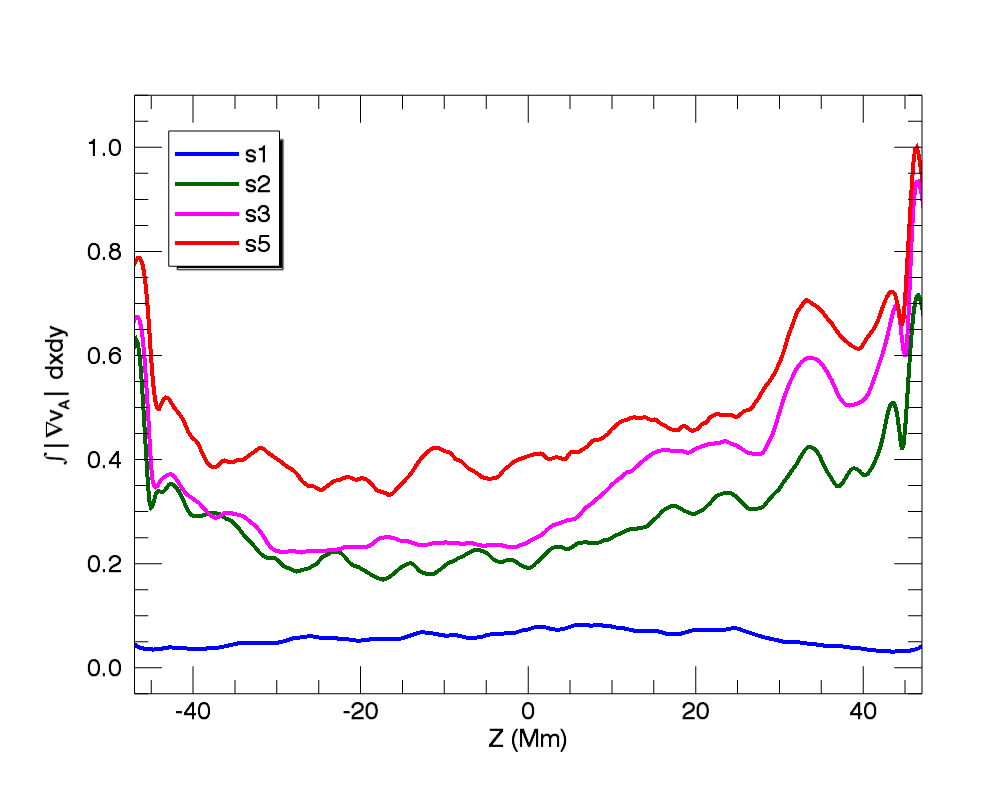}
  \caption{Measure of the field complexity, $I$ (see equation \ref{eq_field_complexity}), in each of the four field configurations as a function of $z$.}
  \label{field_complex_measure}
\end{figure}

In Fig. \ref{field_complex_measure}, we show an additional measure of the small scales within the background field. For each of the four initial conditions (s1, s2, s3 and s5), at each height, $z$, we integrate the magnitude of gradients in the Alfv\'en speed over the $x$-$y$-plane. In each case, we caclulate 
\begin{equation} \label{eq_field_complexity} I(z) = \int_A \lvert \nabla v_A \rvert \, \, \mathrm{d}x \mathrm{d}y, \end{equation} 
and normalise by the maximum of $I(z)$ in the s5 simulation. This quantity acts as a proxy for the rate of deformation (phase mixing) in the wave front as it propagates along magnetic field lines. In Fig. \ref{field_complex_measure}, we see that this measure predicts more significant phase mixing in the simulations with more complex field structures.

In addition to the initial states described above, for the purposes of comparison, we will also consider the behaviour of waves in a uniform domain with a magnetic field aligned with the vertical ($z$) axis. In this case, the initial conditions correspond to the plasma prior to any rotational driving is applied \citep[see][or Table \ref{Tab_eqm_values} for details of plasma parameters]{Reid2018}. 

\begin{table*}[]
\begin{tabular}{c|ccc|ccc|ccc|ccc}
\multirow{2}{*}{Simulation} & \multicolumn{3}{c|}{Temperature (MK)} & \multicolumn{3}{c|}{Field Strength (G)} & \multicolumn{3}{c|}{Density ($\rho_0$)} & \multicolumn{3}{c}{Plasma-$\beta$} \\ %\cline{2-13} 
                  & Min.        & Mean       & Max.       & Min.        & Mean        & Max.        & Min.        & Mean        & Max.        & Min.       & Mean      & Max.      \\ \hline
 & & & & & & & & & & & & \\
Uniform       & 1.9         & 1.9        & 1.9        & 10.0        & 10.0        & 10.0        & 1.0         & 1.0         & 1.0         & 0.13       & 0.13      & 0.13      \\
s1               & 1.9         & 2.0        & 2.4        & 9.7         & 10.0        & 10.4        & 0.9         & 1.0         & 1.1         & 0.12       & 0.14      & 0.16      \\
s2               & 1.9         & 2.1        & 8.7        & 9.5         & 10.0        & 10.8        & 0.3         & 1.0         & 1.3         & 0.13       & 0.14      & 0.23      \\
s3               & 1.9         & 2.2        & 8.3        & 9.3         & 10.0        & 11.0        & 0.4         & 1.0         & 1.3         & 0.12       & 0.15      & 0.24      \\
s5              & 1.9         & 2.4        & 8.3        & 8.9         & 10.0        & 11.7        & 0.4         & 1.0         & 1.4         & 0.12       & 0.16      & 0.27      \\
t5 (High Res.)  &     1.9        &    2.4        &     10.3       &       9.6      &     10.0        &    10.7         &     0.3        &     1.0        &      1.4       &    0.12        &    0.17       &   0.24       
\end{tabular}
\caption{Mean, minimum and maximum values of important plasma and magnetic field parameters for the simulations discussed within this study. The quantities relate to the state of the plasma following the numerical relaxation and prior to the introduction of the wave front.}
\label{Tab_eqm_values}
\end{table*}

\subsection{Boundary conditions}
Following the numerical relaxation, a time-dependent velocity driver was imposed on the lower $z$ boundary in order to introduce a transverse wave into the domain. Hereafter, $t = 0$ refers to the start time of this driving in each simulation. In subsequent sections, we conduct an analysis on different wave drivers (detailed below), however, in the most simple case, the imposed velocity has the form $\vec{v} = (0, v_y(t), 0)$, where   
\begin{equation}\label{wd_eqn1}v_y = \begin{cases}
v_0\sin\left(\frac{2\pi t}{\tau}\right) \hspace{1.5cm} &\text{if } t \le \tau,\\
0  \hspace{1.5cm} &\text{if } t > \tau. 
\end{cases}
\end{equation}
Here, $v_0$ is the amplitude of the driver and $\tau$ is the period of the driver. For the purpose of these simulations, we set $v_0 \approx 20 \text{ km s}^{-1}$ and $\tau \approx 30$ s. We note that the velocity amplitude is much smaller than the local Alfv\'en speed ($ 560 \text{ km s}^{-1} < v_A < 1200 \text{ km s}^{-1}$ with a mean of approximately 700 $\text{km s}^{-1}$) and is similar in magnitude to wave amplitudes observed in the solar atmosphere \citep[e.g.][]{McIntosh2011, Morton2015}. Additionally, we observe that the selected wave period is much shorter than the frequently studied 3-5 minute period which could be excited by p-modes at the photospheric surface \citep[see][for example]{Tomczyk2009, Morton2016}. Such a short period was selected in order to ensure that the height of our domain (100 Mm) will contain several wavelengths of the propagating pulse. Furthermore, we highlight that, in the current form, only a single pulse is introduced into the domain. Numerical simulations implementing continuous wave driving will be the focus of subsequent work. 

\begin{figure}[h]
  \centering
  \includegraphics[width=0.5\textwidth]{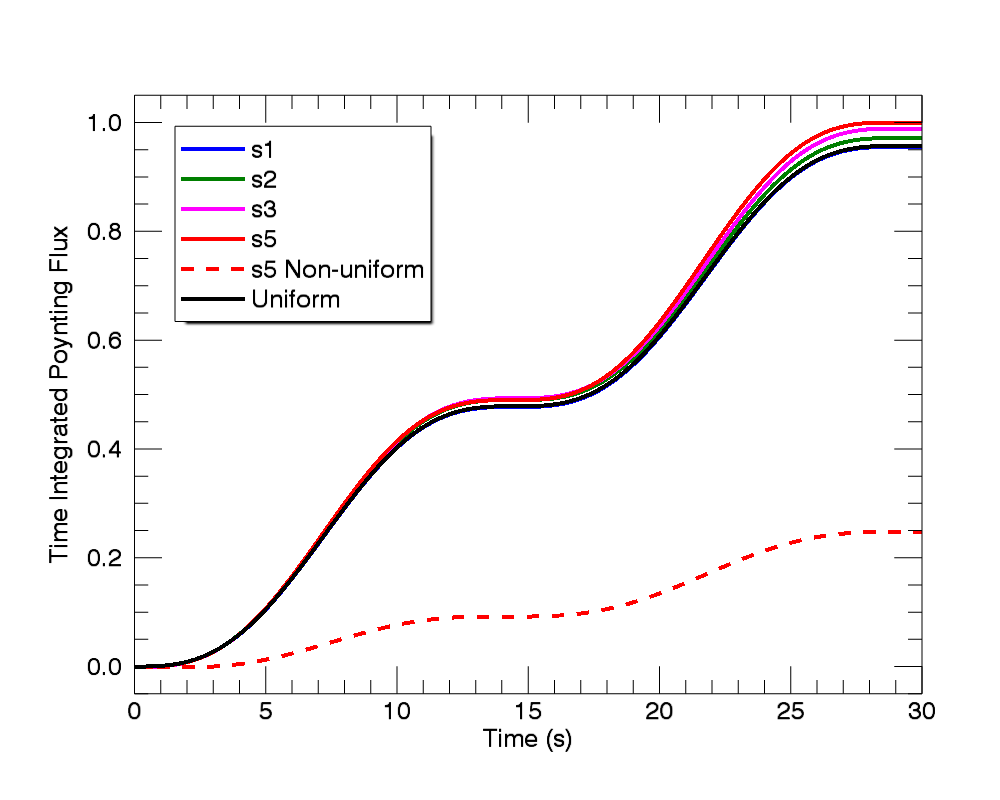}
  \caption{Energy injected into the domain by the wave driver. We display the time integral of the Poynting flux through the $z = -50$ Mm (lower) boundary for different initial conditions with the drivers described by equations \ref{wd_eqn1} and \ref{wd_eqn2} (solid and dashed lines, respectively). Here we have normalised by the maximum of the s5 (solid red) curve.}
  \label{Poynting_injection}
\end{figure}

In Fig. \ref{Poynting_injection}, we display the energy injected by the wave driver described by equation \ref{wd_eqn1} (solid curves) for different initial conditions (s1-s5 and the uniform field case). Since there are no flows through the boundary, this corresponds to the time integral of the Poynting flux. In these simulations, the mean Poynting flux of energy into the domain (during the 30 s wave driving period) is around 50 $\text{W m}^{-2}$. Even if all of the injected energy is dissipated, this is approximately an order of magnitude below the expected energy requirements in the quiet Sun \citep{Withbroe1977}. Indeed, in each case, the total energy increase is much less than 0.1\% of the initial magnetic and thermal energies. 

The small difference between each of the solid curves arises due to the differences in the background field complexities and the weak non-linearity of the wave driver. In particular, for the same velocity driver, a greater Poynting flux is injected for a more complex magnetic field. The dashed red curve corresponds to a modified driver which is described in Section \ref{non-unidriver}.

In all simulations, the $x$ and $y$ boundaries are periodic and field lines are able to cross through the corresponding faces (see, for example, red and purple field lines in the right-hand panel of Fig. \ref{In_Field_Vertical}). For the driven $z$ boundaries, the horizontal velocity components are described above, flows through the boundary are not permitted ($v_z = 0$) and all remaining variables have zero gradients across the boundaries.
%%%%%%%%%%%%%%%%%%%%%%%%%%%%%%%%%%%%%%%%%%%%%%%%%%%%%%%%%%%%%%%%%%%%%%%
%RESULTS
%%%%%%%%%%%%%%%%%%%%%%%%%%%%%%%%%%%%%%%%%%%%%%%%%%%%%%%%%%%%%%%%%%%%%%%
\section{Results}
\subsection{Wave dynamics} \label{Results_section1}
In this section, we analyse the behaviour of the propagating wave in the simulation with the most complex field, the s5 simulation. The wave front introduced by the velocity driver reaches the upper $z$ boundary after approximately $T_e = 145$ s. We select this time, $T_e$, as the end point of the simulation. The mean propagation speed coupled with the selected driver frequency, ensures that the height of the numerical domain (100 Mm) is approximately 5 times the wavelength. 

\begin{figure}
\centering
\begin{subfigure}[b]{0.2\textwidth}
   \includegraphics[width=1\linewidth]{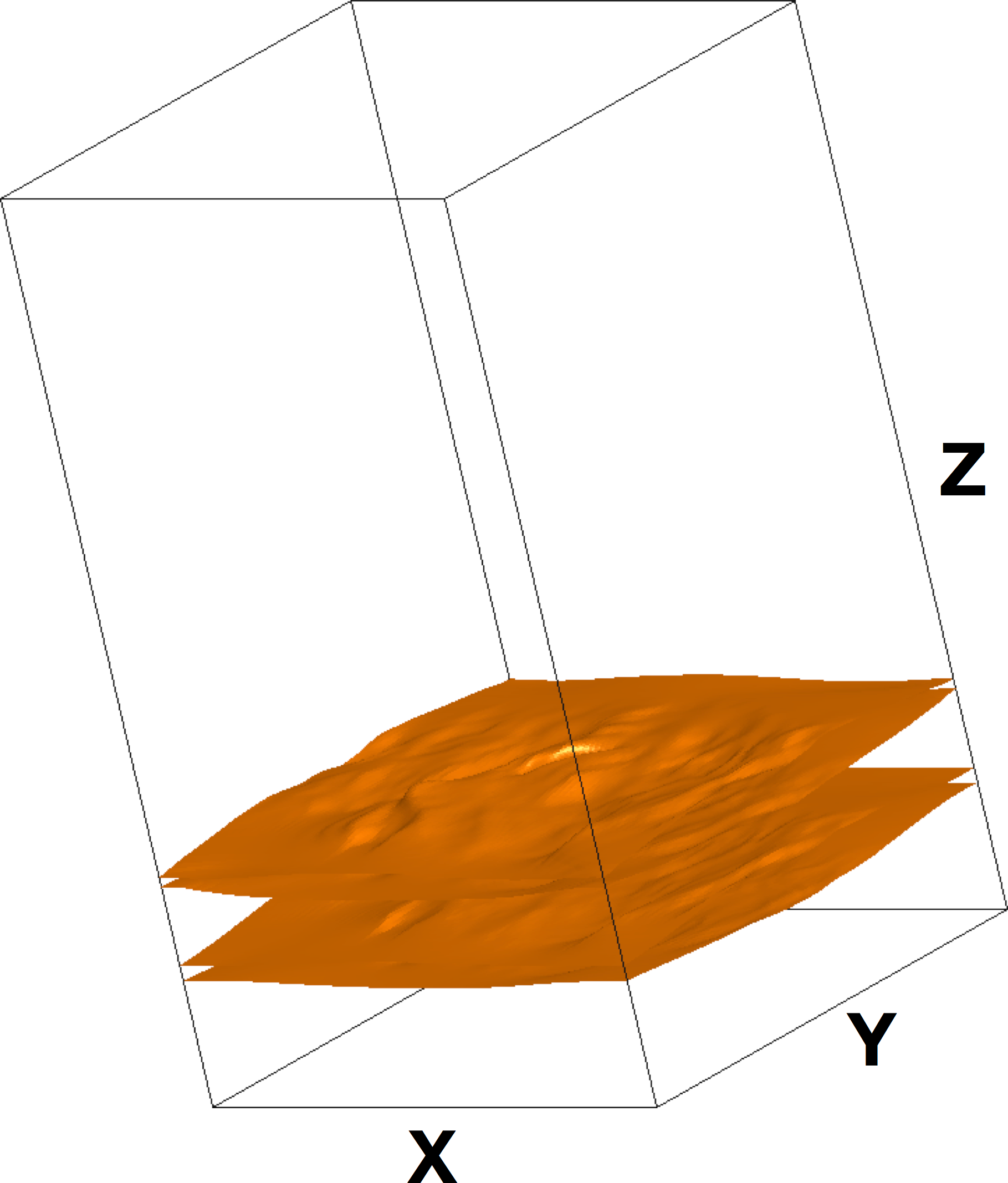}
   \caption{$t = 0.3 \, T_e$.}
   \label{Early_vy_blobs} 
\end{subfigure}
\begin{subfigure}[b]{0.2\textwidth}
   \includegraphics[width=1\linewidth]{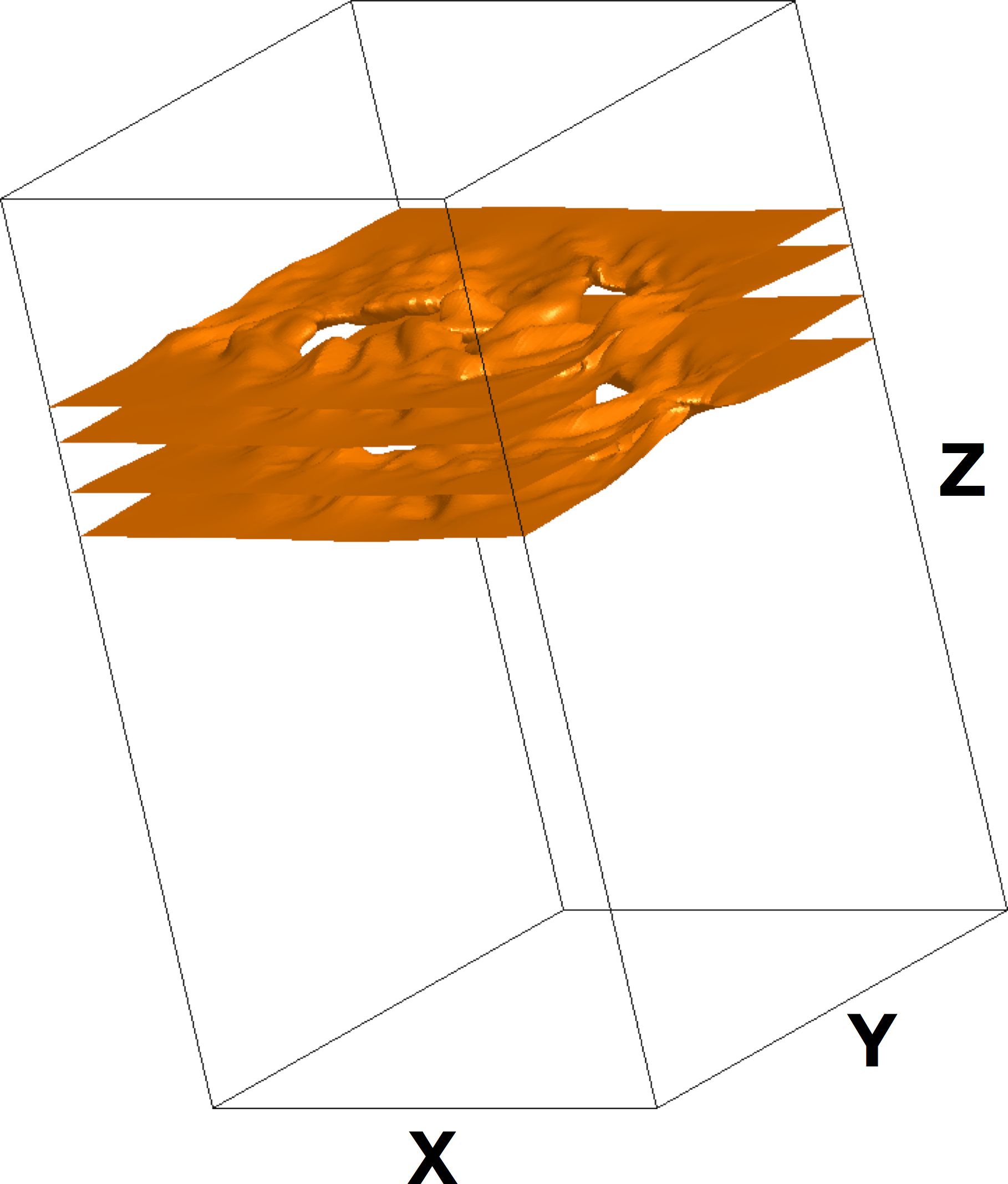}
   \caption{$t = 0.8 \, T_e$.}
   \label{Late_vy_blobs}
\end{subfigure}
\caption{Isosurfaces of the magnitude of $v_y$ corresponding to $\lvert v_y \rvert \approx 0.8 \, v_0$.}
  \label{Vy_blobs}
\end{figure}

Since the wave front propagates through a non-uniform plasma, it cannot be identified as a pure Alfv\'en wave. In particular, it has mixed properties, is mildly compressible and exhibits increasingly complex behaviour as the simulation progresses. In Fig. \ref{Vy_blobs}, we display isosurfaces of the magnitude of $v_y$ at $t=0.3 \, T_e$ (left-hand panel) and $t=0.8 \, T_e$ (right-hand panel). Initially, the wave front is spatially uniform and polarised in the $y$ direction. However, the complex nature of the plasma generates fine spatial variation and transfers energy to the other velocity components. We now describe the two key processes that are associated with the formation of these small spatial scales. 

Firstly, the non-uniform Alfv\'en speed contributes to phase mixing as the wave front propagates at different speeds on neighbouring field lines. Consequently, as time progresses, small length scales develop across the magnetic field. Additionally, the length of field lines vary within the braided region. Therefore, since the wave propagates along magnetic structures, this enhances the rate of phase mixing and deformation of the wave packet.

In the classic case of phase mixing \citep[as described in][]{Heyvaerts1983}, magnetic field lines have a constant Alfv\'en speed along their length, ensuring that a wave front will always become increasingly out of phase as time progresses. However, in the current case, the Alfv\'en speed varies along field lines and thus phase mixing does not proceed as outlined in \citet{Heyvaerts1983}. Indeed, it is possible for a phase mixed Alfv\'en wave to become more in phase if it encounters an appropriate Alfv\'en speed profile. This continuous phase mixing and {\emph{reverse phase mixing}} ensures that the wave front does not become as distorted as in the classic case \citep[see the right-hand panel of Fig. 5 in][for example]{Pascoe2010}. However, in the current paradigm, the spatial distribution of gradients in the Alfv\'en speed ensures that the phase mixing is spread across a large cross-sectional area of the wave front.

\begin{figure}
\centering
\begin{subfigure}[b]{0.2\textwidth}
   \includegraphics[width=1\linewidth]{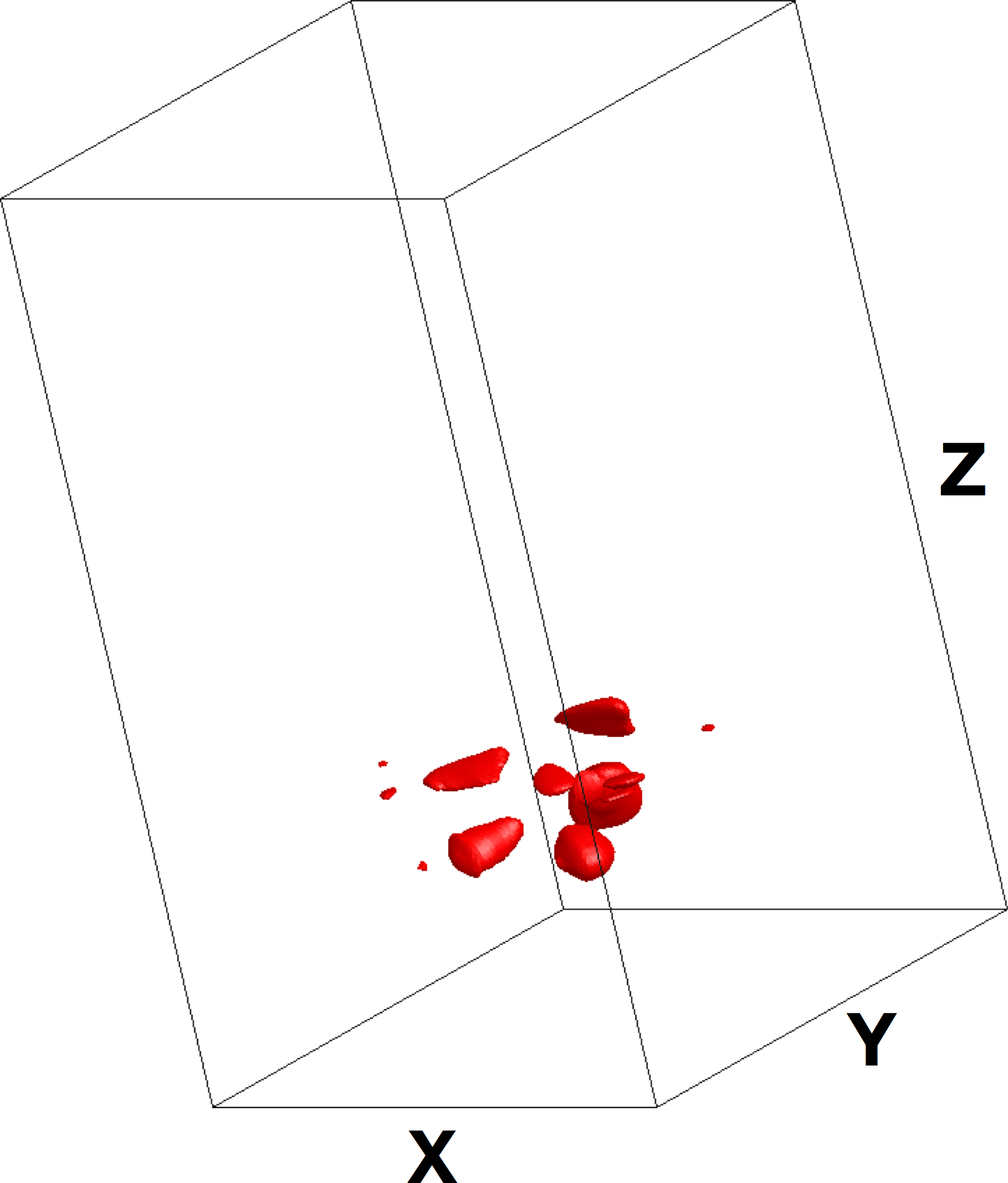}
   \caption{$t = 0.3 \, T_e$.}
   \label{Early_vx_blobs} 
\end{subfigure}
\begin{subfigure}[b]{0.2\textwidth}
   \includegraphics[width=1\linewidth]{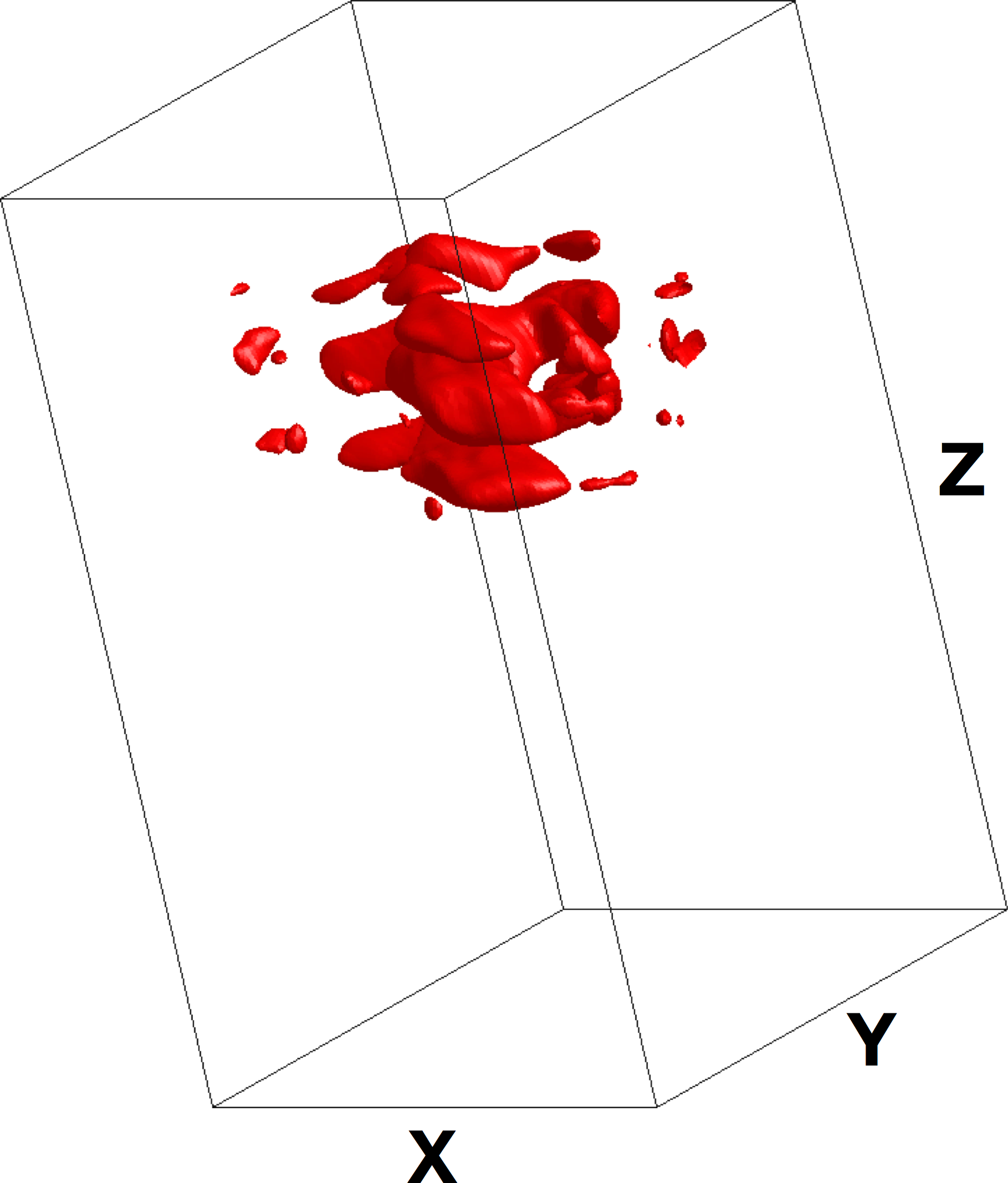}
   \caption{$t = 0.8 \, T_e$.}
   \label{Late_vx_blobs}
\end{subfigure}
\caption{Isosurfaces of the magnitude of $v_x$ corresponding to $\lvert v_x \rvert \approx 0.25 \, v_0$.}
  \label{Vx_blobs}
\end{figure}

Secondly, the rotational nature of the driver imposed in \citet{Reid2018}, ensures that, at $t=0$, in each non-uniform simulation the currents are predominantly aligned in the $z$ direction (see previous section). Meanwhile, the velocity driver induces a perturbation of the $y$ component of the magnetic field. The interaction between this perturbation and the $z$ component of the background currents generates a Lorentz force acting parallel to the $x$ axis. This transfers energy from the initially driven component of the velocity ($v_y$) to the other horizontal component ($v_x$). This has the effect of locally modifying the polarisation of the wave front. There is also a small transfer of energy to $v_z$ associated with horizontal background currents. However in the phase-mixing region the mean of $\lvert v_z\rvert$ is approximately $2\%$ of the mean of $\lvert v_y\rvert$, whereas the mean of $\lvert v_x\rvert$ is approximately $20\%$ of the mean of $\lvert v_y\rvert$.

In Fig. \ref{Vx_blobs}, we show isosurfaces of the magnitude of $v_x$ at $t=0.3 \, T_e$ and $t=0.8 \, T_e$. We notice that at both times the spatial extent of the energy conversion is limited to the regions of most complex field (compare to the lower right-hand panel of Fig \ref{In_Field_Top}, for example) as these are the locations containing the most significant background currents. Further, as time progresses, energy is continuously transferred to the $x$ component of the velocity, resulting in the increase in volume within the isosurface observed between the two panels in Fig. \ref{Vx_blobs}. 

\begin{figure}
\centering
   \includegraphics[width=1\linewidth]{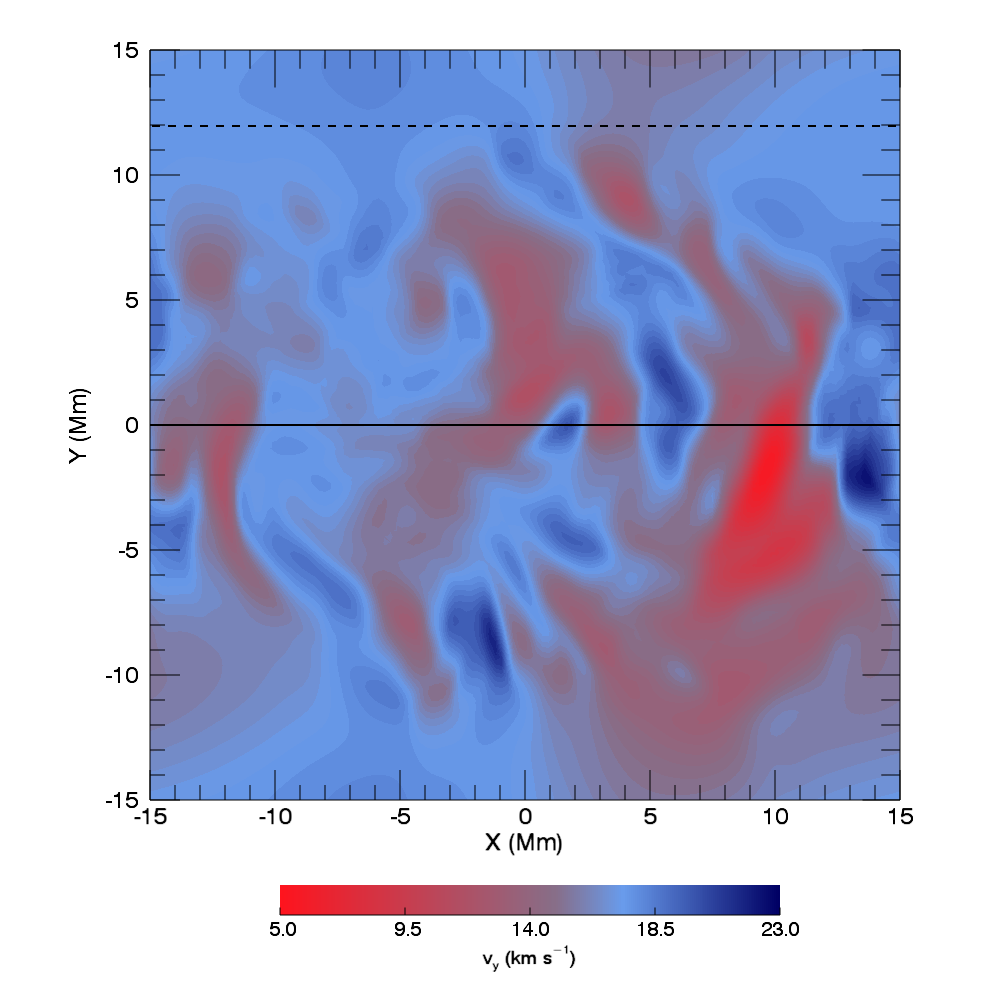}
\caption{The $y$ component of the velocity field in a horizontal cut through the wave front at $t = 0.8 \, T_e$.}
  \label{v_contour}
\end{figure}

\begin{figure}[h]
  \centering
  \includegraphics[width=0.5\textwidth]{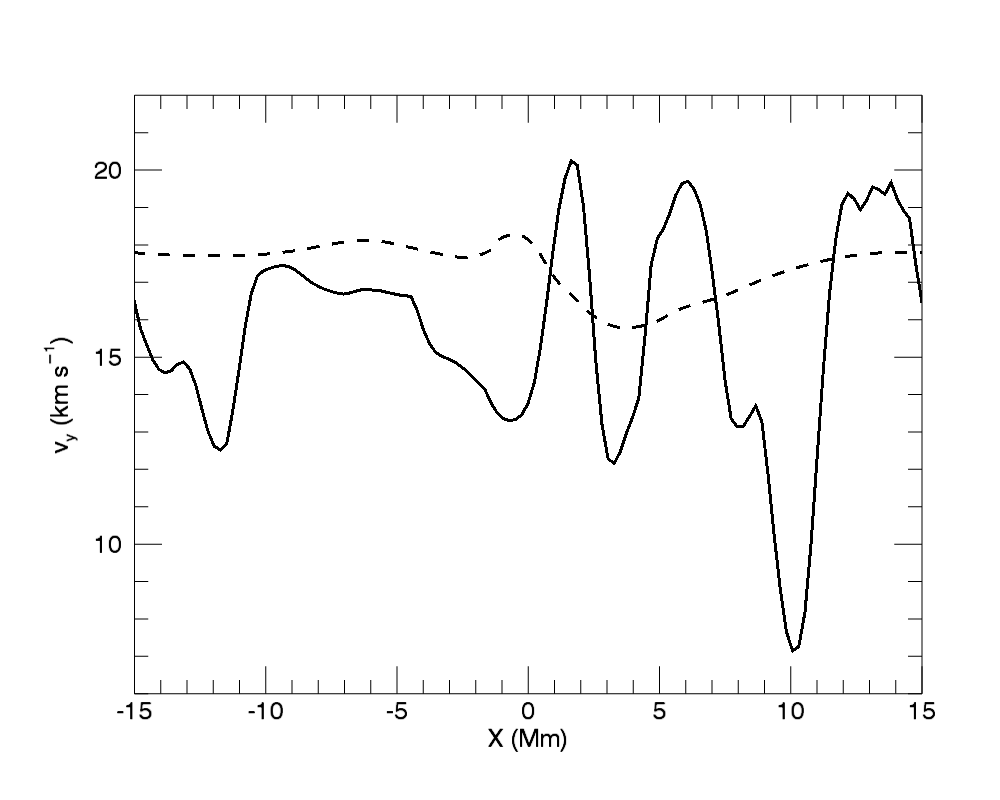}
  \caption{The $y$ component of the velocity field along the two horizontal lines shown in Fig. \ref{v_contour}.}
  \label{v_y_phasemix}
\end{figure}

The combination of these two, highly localised, effects results in the formation of transverse gradients in the perturbed velocity field. The associated small scales are shown in greater detail in Figs. \ref{v_contour} and \ref{v_y_phasemix}. In Fig. \ref{v_contour}, we show the profile of $v_y$ in a horizontal cut through the wave packet at $t = 0.8 \, T_e$. We note that there is spatial variation throughout the entire cross-section but the perturbation is more homogeneous at large $|y|$, where the magnetic field is less complex. The solid and dashed lines shown in Fig. \ref{v_contour} correspond to the locations of the respective lines depicted in Fig. \ref{v_y_phasemix}. Here, we clearly see the large transverse gradients that form as a result of the complexity of the background plasma. In Fig. \ref{v_y_phasemix}, we again highlight the relative homogeneity of the dashed line (region of less complex field) in comparison to the solid line (region of more complex field).

\begin{figure}[h]
  \centering
  \includegraphics[width=0.5\textwidth]{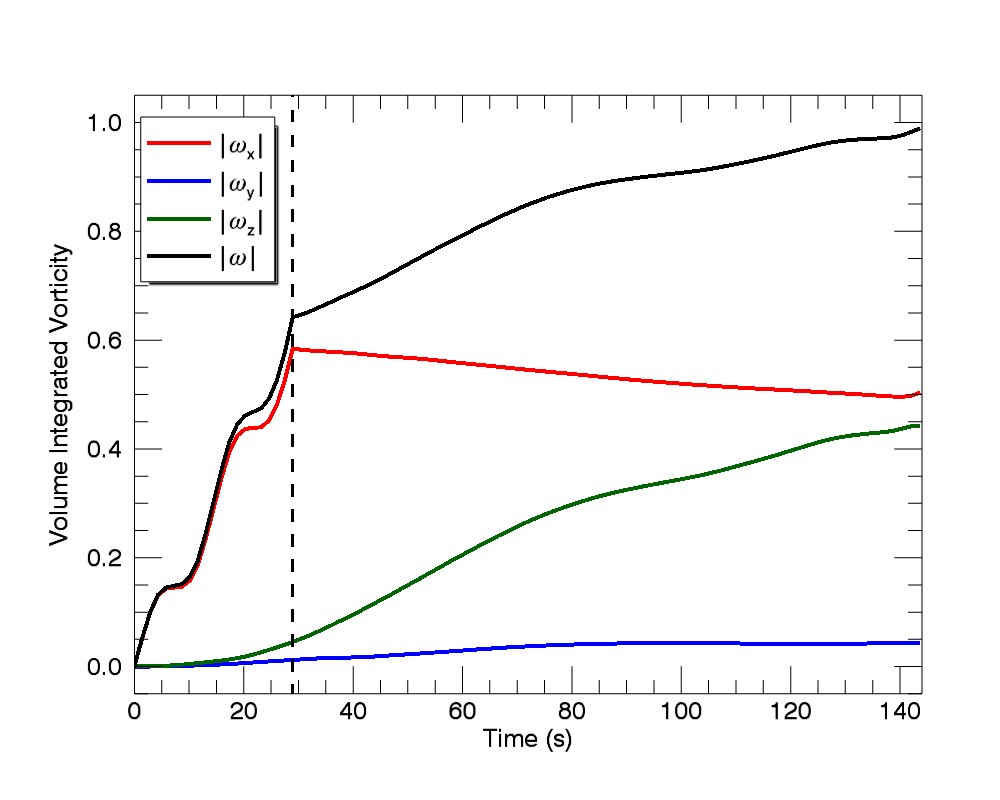}
  \caption{Volume integrated $\lvert \omega \rvert$ as a function of time for simulation, s5. Additionally, we display the volume integral of the magnitude of each component and the dashed black line corresponds to the time $t = \tau$ when the wave driving ceases.}
  \label{vort_comp}
\end{figure}

The formation of these small spatial scales in the velocity field can be tracked using the vorticity, ${\vec{\omega}} = \nabla \times \vec{v}$. The imposed wave driver injects vorticity as it introduces vertical gradients in the driven component of the velocity $\left(\dfrac{\partial v_y}{\partial z}\right)$, and hence contributes to an increase in the $\omega_x$ term. Meanwhile, the formation of small scales caused by the background field complexity contributes to the formation of transverse gradients of $v_y$ and to a lesser extent $v_x$. This results in a gradual increase in the magnitude of $\omega_z$. 

We display this behaviour by showing the change in the volume integrated magnitude of vorticity and of each component in Fig. \ref{vort_comp} as the simulation progresses. We see an initial increase in $\omega_x$ (red line) as the driver introduces a wave pulse into the domain. However, at $t \approx 29$ s, the driving motion is suspended and the magnitude of $\omega_x$ component slowly decreases as energy is transferred from the $y$ component of the velocity (see above). Meanwhile, there is a steady increase in the $z$ component of vorticity (green line) as phase mixing progresses. This increase is much greater than the decrease observed in $\omega_x$, resulting in a net increase in total vorticity (black line) during the simulation. This will produce an enhancement in the rate of viscous dissipation as we shall discuss later. We note that there is also a small increase in the volume integral of $\lvert\omega_y\rvert$ (blue line) as time progresses. This is predominantly associated with the localised formation of vertical gradients in $v_x$ (see Fig. \ref{Vx_blobs}).

\begin{figure}
\centering
\begin{subfigure}[b]{0.2\textwidth}
   \includegraphics[width=1\linewidth]{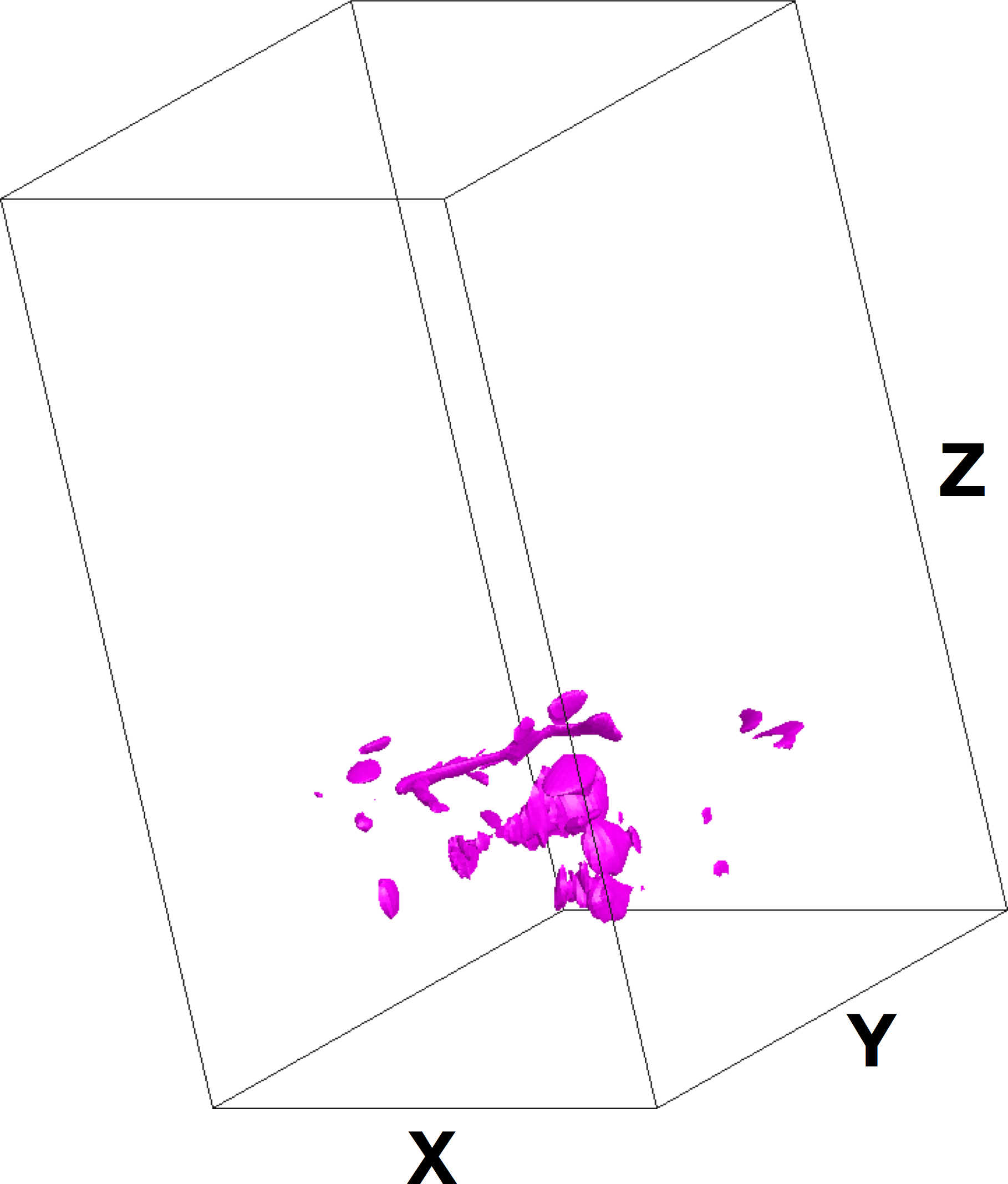}
   \caption{$t = 0.3 \, T_e$.}
   \label{Early_vort_blobs} 
\end{subfigure}
\begin{subfigure}[b]{0.2\textwidth}
   \includegraphics[width=1\linewidth]{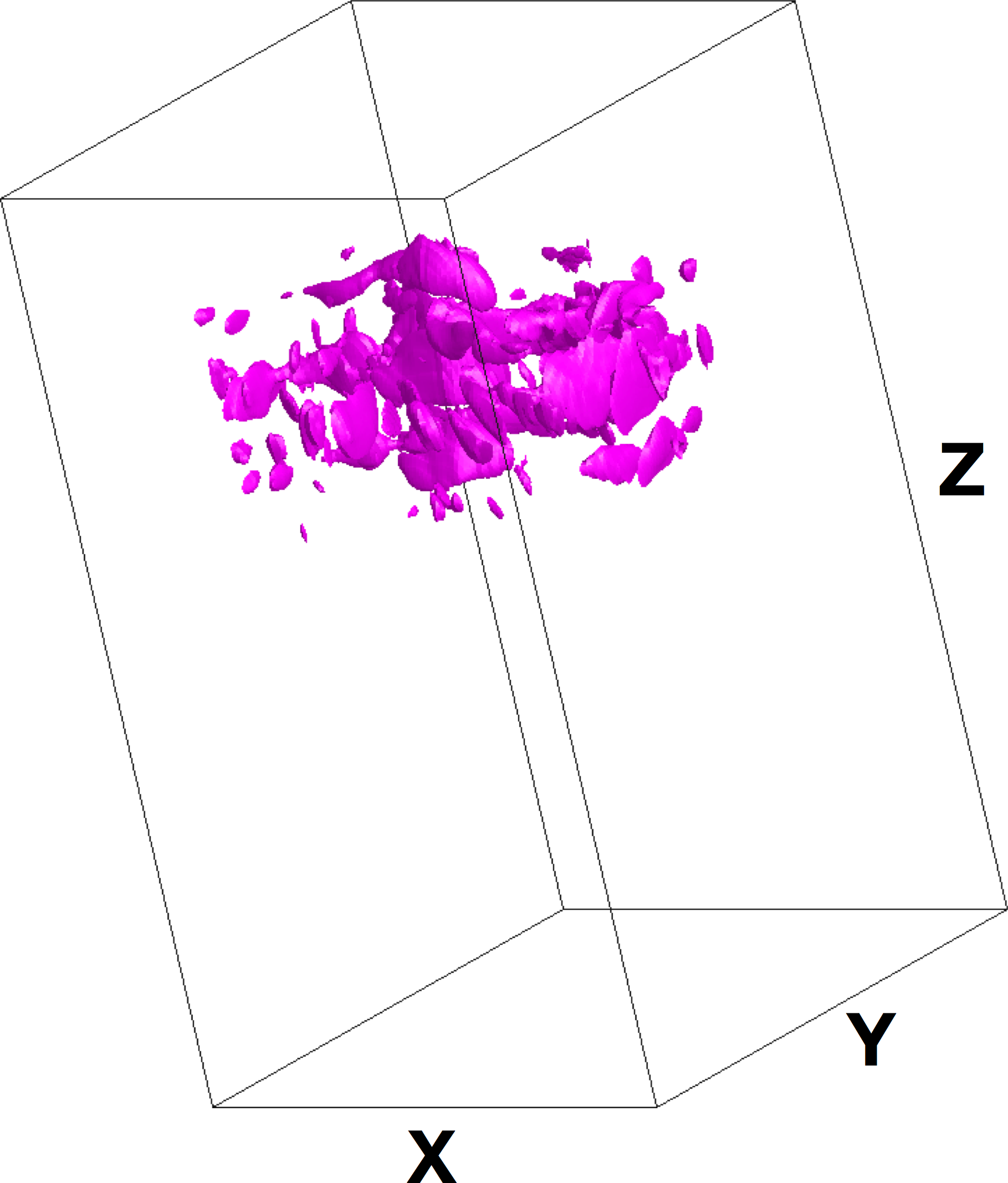}
   \caption{$t = 0.8 \, T_e$.}
   \label{Late_vort_blobs}
\end{subfigure}
\caption{Isosurfaces of the magnitude of $\vec{\omega}$ in the s5 simulation.}
  \label{Vorticity_blobs}
\end{figure}

In Fig. \ref{Vorticity_blobs}, we display an isosurface of $\lvert\vec{\omega}\rvert$ at $t=0.3 \, T_e$ (left-hand panel) and $t=0.8 \, T_e$ (right-hand panel) for the s5 simulation. At both times, we see that the formation of large vorticities is confined to regions of complex field and even within this sub-section of the domain, the vorticity has a fragmented nature. This suggests that any viscous heating associated with this wave will not heat uniformally. Despite this, in contrast with many classical phase mixing simulations, small spatial scales form over a larger proportion of the cross-sectional area of the magnetic structure and not simply in a small region in the boundary of a coronal loop. Thus, in a non-ideal regime, viscous heating will deposit wave energy throughout the complex field. We will consider the rate of wave energy dissipation in a subsequent section.            

At this stage, it is important to note that the wave dynamics are also associated with the development of small scales in the magnetic field. As such, for comparable values of resistivity, $\eta$, and viscosity, $\nu$, Ohmic and viscous heating are expected to be equally effective as mechanisms for dissipating wave energy. However, since the background field is already finely structured, it is difficult to separate the wave-induced small scales in the magnetic field from the pre-existing gradients. Therefore, for simplicity, we have restricted our analysis to the evolution of small scales in the perturbed velocity field.

\subsection{Field complexity}
\begin{figure}
\centering
\begin{subfigure}[b]{0.2\textwidth}
   \includegraphics[width=1\linewidth]{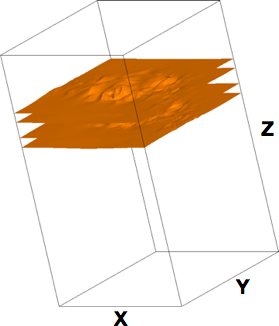}
   \caption{s2.}
   \label{s4vyblobs} 
\end{subfigure}
\begin{subfigure}[b]{0.2\textwidth}
   \includegraphics[width=1\linewidth]{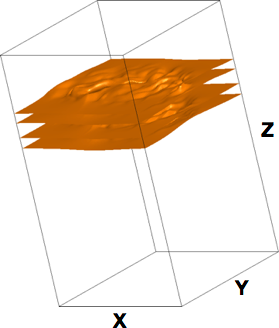}
   \caption{s3.}
   \label{s6vyblobs}
\end{subfigure}
\caption{Isosurfaces of the magnitude of $v_y$ corresponding to $\lvert v_y \rvert \approx 0.8 \, v_0$ at $t=0.8 \, T_e$.}
  \label{Vy_blobs2}
\end{figure}

\begin{figure}
\centering
\begin{subfigure}[b]{0.2\textwidth}
   \includegraphics[width=1\linewidth]{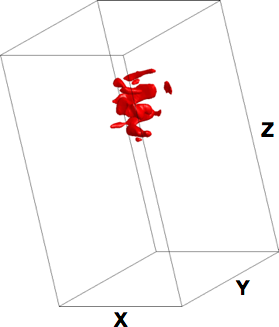}
   \caption{s2.}
   \label{s4vxblobs} 
\end{subfigure}
\begin{subfigure}[b]{0.2\textwidth}
   \includegraphics[width=1\linewidth]{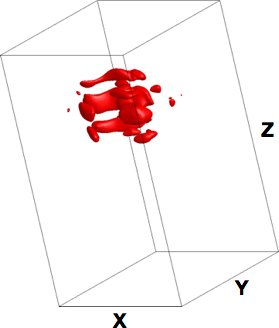}
   \caption{s3.}
   \label{s6vxblobs}
\end{subfigure}
\caption{Isosurfaces of the magnitude of $v_x$ corresponding to $\lvert v_x \rvert \approx 0.25 \, v_0$ at $t=0.8 \, T_e$.}
  \label{Vx_blobs2}
\end{figure}

In order to quantify the effects of varying the field complexity, we now contrast the observed wave dynamics in the s1, s2, s3 and s5 simulations. In Figs. \ref{Vy_blobs2} and \ref{Vx_blobs2}, we show isosurfaces of $\lvert v_y \rvert$ and $\lvert v_x \rvert$, respectively at $t=0.8 \, T_e$ for the s2 (left-hand panels) and s3 (right-hand panels) simulations. These correspond to Figs. \ref{Late_vy_blobs} and \ref{Late_vx_blobs} which show the corresponding s5 results and were discussed above. Similar plots for the s1 simulation show an almost perfectly planar wave in the $v_y$ component, and very little transfer of energy to the $v_x$ component. From this collection of plots we see that the rate of phase mixing and, thus deformation of the propagating wave front, is sensitive to the complexity of the magnetic field and background Alfv\'en speed profile. 

\begin{figure}[h]
  \centering
  \includegraphics[width=0.5\textwidth]{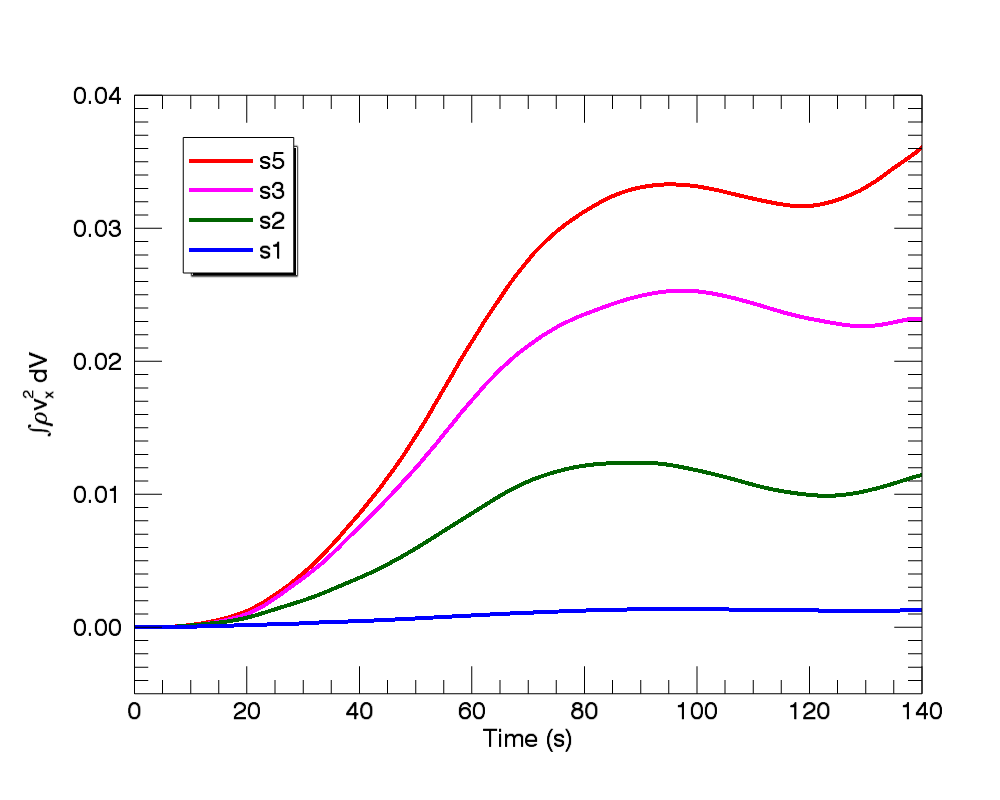}
  \caption{Time evolution of the volume integrated $x$ component of the kinetic energy, $K_x$ (see equation \ref{eq_x_kin_en}), for each of the four complex field simulations.}
  \label{Kinetic_energy_x}
\end{figure}

\begin{figure*}[h]
  \centering
  \includegraphics[width=\textwidth]{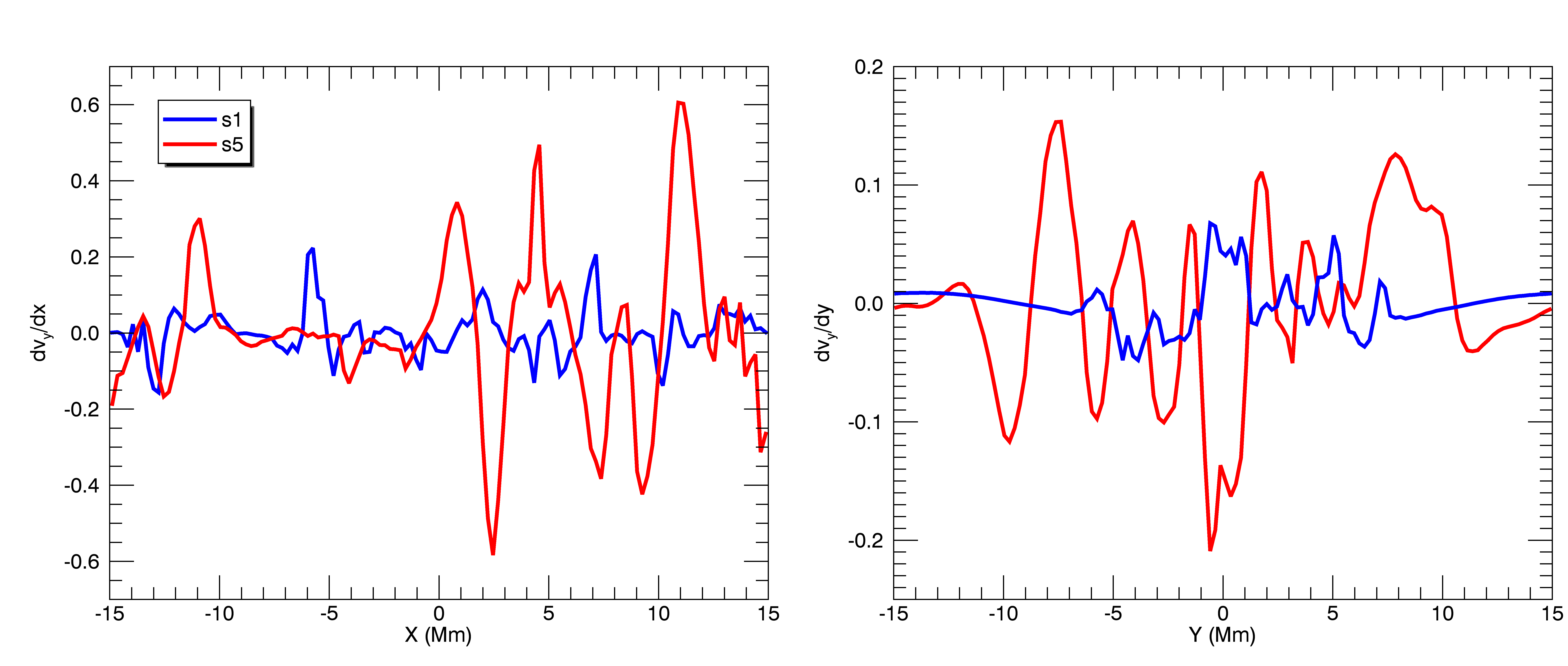}
  \caption{Transverse gradients of $v_y$ for the s1 (blue) and s5 (red) simulations. We show $\dfrac{dv_y}{dx}$ along $y = 0$ Mm (left-hand panel) and $\dfrac{dv_y}{dy}$ along $x = 0 $ Mm (right-hand panel) at $t=0.8\,T_e$ through the location of the wave front. In each case we have normalised by the maximum transverse gradient of $v_y$ in the horizontal plane containing these lines.}
  \label{phase_mix_fc_compare}
\end{figure*}

In particular, we observe that the isosurfaces of $\lvert v_y \rvert$ exhibit greater variation in the more complex field simulations. Not only are the gradients larger (see Fig. \ref{phase_mix_fc_compare} and associated discussion for more detail), but they also occur over a greater cross-section of the propagating wave front. This is a result of the expansion of the braided field region between the simpler and more complex cases (see Fig. \ref{In_Field_Top} for example). Additionally, the transfer of energy to the $v_x$ component is more efficient in the s3 and s5 simulations than in the simpler cases. Again, the increased cross-sectional area of stressed field translates to the formation of a significant $v_x$ component, and hence modification of wave polarisation, over a larger proportion of the domain.   

In Fig. \ref{Kinetic_energy_x}, we aim to quantify the rate of energy transfer from the $y$ component of the velocity to the $x$ component. We calculate 
\begin{equation}\label{eq_x_kin_en}
K_x = \int_V \rho v_x^2 \, \, \mathrm{d} V,
\end{equation}
the $x$ component of the kinetic energy integrated over the volume of the domain, $V$. We show the time evolution of this quantity for each of the four complex field configurations. Meanwhile, the uniform field does not induce any transfer of energy between the transverse velocity components. In Fig. \ref{Kinetic_energy_x}, we have normalised by the total kinetic energy injected into the domain by the wave driver. As such, we see that only a relatively small (of the order 4\% for the most complex field simulation) proportion of the energy is transfered between the $v_y$ and $v_x$ components. However, this is a highly localised process and in some regions of the wave front, the magnitude of $v_x$ is comparable to the size of $v_y$. The observational significance of this result will be examined in subsequent work. 

We discussed the cause of this energy transfer previously and will simply note that the larger the background currents, the more readily the wave polarisation is modified. Additionally, we note that each simulation will induce small perturbations in the field aligned component due to the non-linear ponderomotive force, which will contribute to the $x$ component of the kinetic energy in regions of non-zero $B_x$. Since the strength of the perturbed magnetic field is small in comparison to the background field, we expect this to be a relatively weak effect.   

In Fig. \ref{phase_mix_fc_compare}, we compare the phase mixing gradients that form during the course of the s1 (blue) and s5 (red) simulations. We show $\dfrac{dv_y}{dx}$ and $\dfrac{dv_y}{dy}$ along horizontal lines through the wave front in the left-hand and right-hand panels, respectively. The term shown in the left-hand panel does contribute to the growth in the magnitude of the vorticity (discussed above). The right-hand panel, on the other hand, depicts a term that is not directly tracked by the vorticity. Instead, this contributes to the compression of the plasma and hence may be associated with an observational signature that is absent from a pure, incompressible Alfv\'en wave. Whilst detailed analysis of this effect remains beyond the scope of this work, we note that the plasma compression observed in these simulations is highly inhomogeneous and may provide seismological information about the nature of the background magnetic field.

In both cases, it is clear that the larger gradients (both positive and negative) form in the more complex field simulation (red curves). This is indicative of the enhanced phase mixing observed in the s5 simulation. Further, we note that the magnitude of the largest gradients in the left-hand panel are greater than those in the right-hand panel (see different scales for $y$-axes). This is because the compression of plasma and magnetic field associated with the right-hand panel generates a total pressure force which limits the formation of very small scales in the direction of the velocity perturbation. 

Additionally, we highlight the relatively small gradients in the right-hand panel for large $\lvert y \rvert$. These are particularly apparent in the s1 simulation (blue curve) and are a result of the relatively simply field and Alfv\'en speed profile present in this region of the domain. As such we see very little phase mixing and plasma compression at large values of $\lvert y \rvert$. Thus, by comparing the (potentially observable) plasma compression profiles, we can deduce information about the relative complexity of the field at different values of $\lvert y \rvert$. The observational implications of this effect will be explored in future work.

\label{Diss_sec}
\begin{figure}[h]
  \centering
  \includegraphics[width=0.5\textwidth]{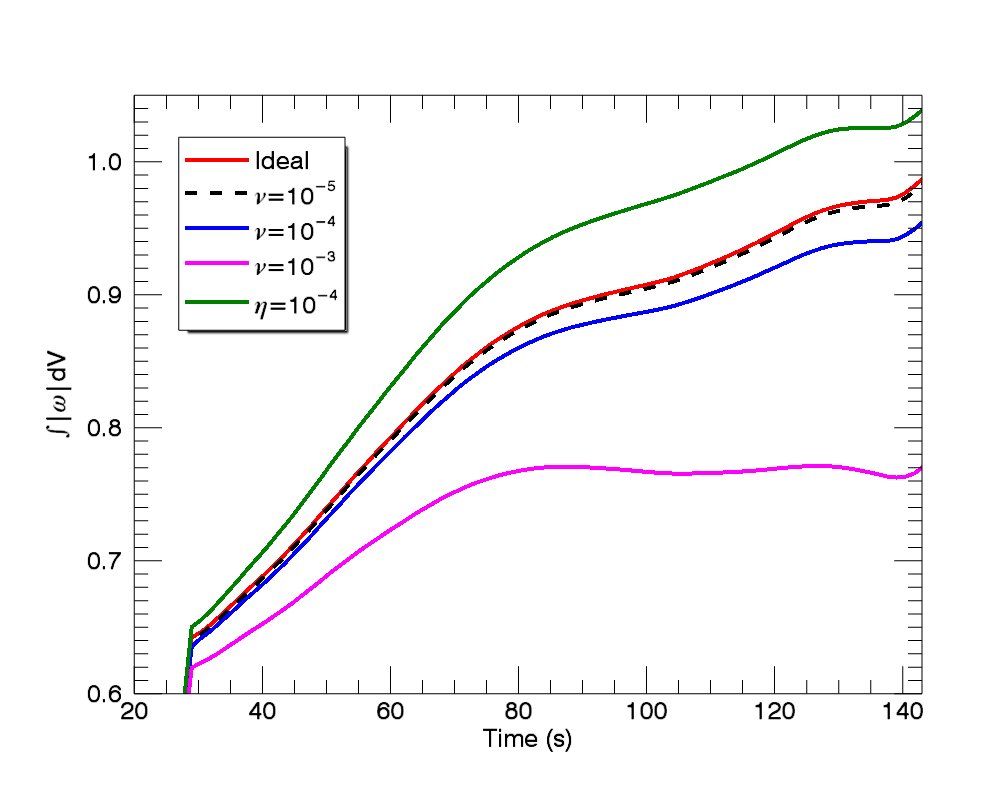}
  \caption{Time evolution of $\lvert\omega \rvert$ integrated over the numerical domain for the s5 simulation with different transport coefficients. We note that the early stages of the simulation are omitted here and we have normalised the integral using the maximum value attained in the ideal s5 simulation (as in Fig. \ref{vort_comp}).}
  \label{vort_diss_comp}
\end{figure}

\subsection{Dissipation}

\begin{figure*}[h]
  \centering
  \includegraphics[width=\textwidth]{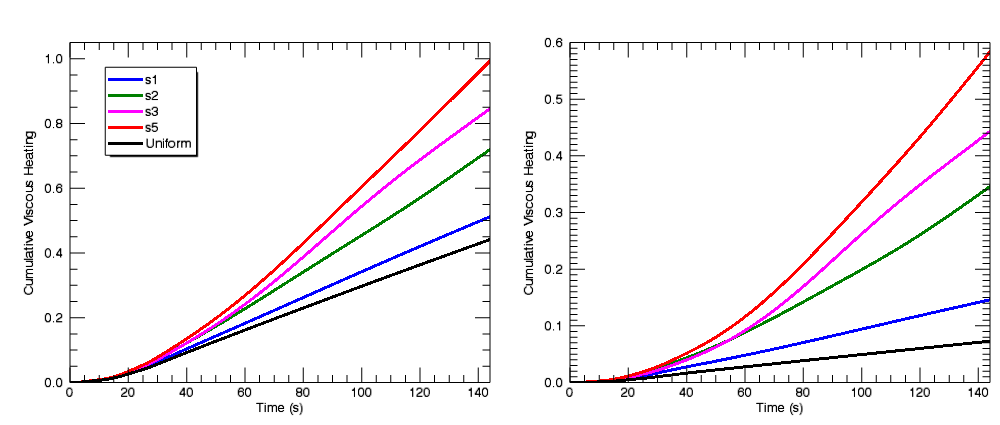}
  \caption{Cumulative volume integrated viscous heating for the braided and uniform field simulations. We show the heating for $\nu=10^{-3}$ in the left-hand panel and for $\nu=10^{-4}$ in the right-hand panel. In each case we have normalised by the total heating obseverd in the s5 simulation with $\nu=10^{-3}$.}
  \label{visc_heating_compare}
\end{figure*}

In this section, we seek to quantify the significance of the previously discussed formation of small spatial scales in relation to the dissipation of wave energy. In particular, we consider the effects of non-zero transport coefficients on the evolution of the simulations. As a non-zero resistivity, $\eta$, will cause the dissipation of energy in the non-potential background field (in addition to any wave heating), we predominantly focus on the effects of viscosity, $\nu$. With this approach, we do not expect significant energy to be extracted from the background conditions.

We repeat the previously described simulations with values of viscosity, $\nu = 10 ^{-3}, 10^{-4}$ and $\nu = 10^{-5}$. These correspond to Reynolds numbers of $10^{3}, 10^{4}$ and $10^{5}$, respectively, and act in addition to the aforementioned shock viscosities (see Section 2) which also provide dissipative effects. 

In Fig. \ref{vort_diss_comp}, we display the volume integral of the magnitude of the vorticity (also see Fig. \ref{vort_comp}) for the s5 simulation with various values of $\nu$. For comparison, we also include the results of a simulation with $\eta = 10^{-4}, \nu=0$ (green curve). We see that the formation of small scales in the velocity field is suppressed by increasing the frictional effects of viscosity (larger values of $\nu$). The similarity between the red and black dashed curves confirms that at $\nu=10^{-5}$, numerical dissipation and the effects of the shock viscosities have greater significance for the wave dynamics. As such, the maximum Reynolds number obtained with the current numerical resolution in these simulations is of the order $10^{4} - 10^{5}$.

Counterintuitively, increasing the resistivity enhances the total vorticity observed within the domain (compare green and red curves). This is not a result of the wave dynamics but instead is caused by the diffusion and Ohmic dissipation of the background magnetic field. This process generates small, localised flows and contributes to the increase of the vorticity. The phase mixing of the wave front is still suppressed in this case, however the associated decrease in small scale formation is dominated by the effects caused by the non-zero $\eta$ acting on the background field.

In Fig. \ref{visc_heating_compare}, we display the cumulative volume integrated viscous heating for the s1, s2, s3, s5 and uniform simulations for $\nu = 10^{-3}$ (left-hand panel) and $\nu = 10^{-4}$ (right-hand panel). This heating is associated with these values of $\nu$ and the shock viscosities which are not changed between simulations. In the Lare3d code, numerical dissipation does not contribute to the heating of plasma. It is important to note that due to the nature of the wave driver (low amplitude and only a single pulse), very little energy is injected into the domain. As such, we cannot expect significant plasma heating to occur in this case (see Sect. \ref{Sect_Discussion} for further discussion). However, continuous driving may provide sufficient energy in wave heating models \citep[e.g.][]{Karampelas2018} and this possibility will be investigated in future studies. 

In the left-hand panel of Fig. \ref{visc_heating_compare}, we see that the viscous disspation of wave energy is a little more than twice as efficient in the s5 simulation (red) line than in the uniform field case (black line). Indeed, we see that increasing the field complexity gradually increases the total viscous heating. Despite this, given that we expect very little heating from a planar Alfv\'en wave in the high Reynolds number corona, the increase in energy dissipation due to phase mixing seems insufficient to generate significant wave heating. However, in the right-hand panel we see that the relative increase of viscous heating between the uniform and most braided field cases is enhanced at lower values of $\nu$. This is because the formation of small scales due to phase mixing is more efficient at lower viscosities for the s5 simulation (no phase mixing occurs in the uniform field case). 

By comparing the two panels of Fig. \ref{visc_heating_compare}, we see that the viscous heating does not simply scale with the size of $\nu$. In particular, the energy dissipation depicted in the right-hand panel is not an order of magnitude smaller than in the left-hand panel. There are two main reasons for this. Firstly, the formation of small scales in the velocity field is more efficient in the higher Reynolds number case (right-hand panel) and thus the viscosity is acting on larger spatial gradients. Secondly, the magnitudes of the shock viscosities are unchanged between the two sets of simulations and thus the effects of these terms are more significant when the \emph{real} viscosity is reduced.   

\subsection{Resolution considerations}
In the model presented within \citet{Reid2018}, the widths of the current sheets that form are a function of the spatial resolution implemented in the numerical code. In particular, higher resolution simulations allow the formation of narrower current sheets and smaller scale inhomogeneties in the plasma as the numerical dissipation is reduced. In this section we consider a new initial state, t5, which is obtained using a domain with $256 \times 256 \times 1024$ grid cells. In this case, the same uniform field is driven with identical rotational motions for 500 Alfv\'en times. The plasma is then allowed to relax numerically under the effects of a high viscosity as with all previously discussed simulations. For comparison with the lower resolution experiments, important characteristics of the initial state are included in Table \ref{Tab_eqm_values}. Once a numerical equilibrium is obtained, an identical, single pulse sinusoidal wave is intoduced at the $z=-50$ Mm (lower) boundary. 

\begin{figure}[h]
  \centering
  \includegraphics[width=0.5\textwidth]{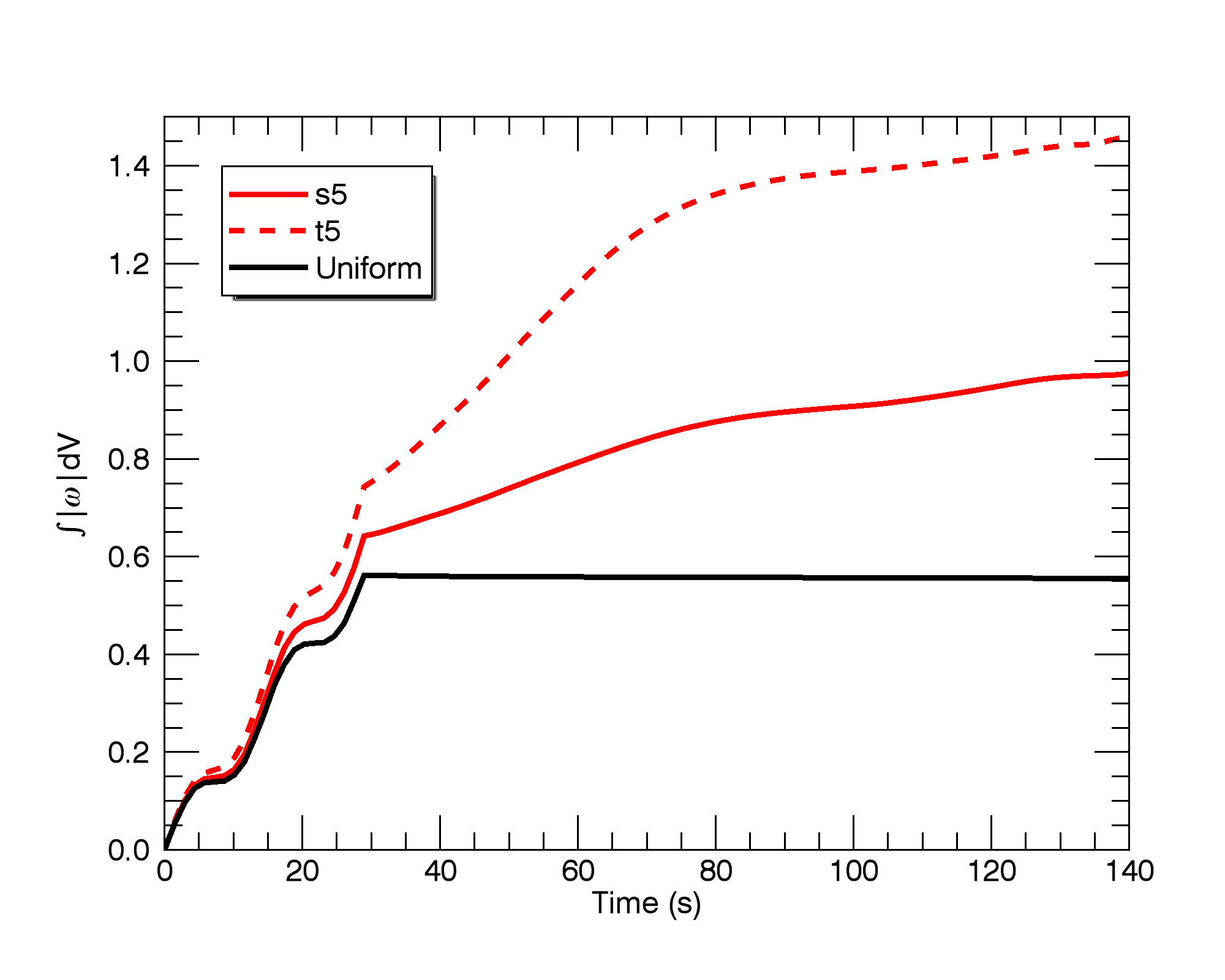}
  \caption{Volume integrated $\lvert \omega \rvert$ as a function of time for the s5, t5 and uniform field simulations. In each case, we have normalised by the maximum value of the integral for the s5 simulation (see also Figs. \ref{vort_comp} and \ref{vort_diss_comp}).}
  \label{vort_res_compare}
\end{figure}

In the higher resolution simulation, the fundamental wave dynamics remain unchanged, however, the increased intricacy of the initial conditions induce enhanced phase mixing and allow even smaller spatial scales to form. In Fig. \ref{vort_res_compare}, we quantify this effect by plotting the time evolution of the volume integral of the magnitude of the vorticity for the s5 (solid red), t5 (dashed red) and the uniform field (black) simulations. As in Fig. \ref{vort_comp}, we observe an initial rise in the vorticity in all cases as a result of the gradients introduced by the wave driver. Subsequently, the progression of phase mixing in the complex field simulations (but not in the uniform case) enhances the volume integrated vorticity. 

The vorticities in the uniform field case represent the vertical gradients of $v_y$ associated with the Alfv\'en wave. As such, they are inversely proportional to the wavelength of the perturbation. The red curves are a combination of these gradients and the small scales associated with phase mixing. The increase observed in the dashed red line in comparison to the solid red line is simply due to the enhanced phase mixing observed in the high resolution simulation. 

The increased field line and Alfv\'en speed complexity causes a greater deformation in the propagating wave front. Additionally, since the effects of numerical dissipation are reduced in the t5 simulation, larger gradients in the velocity field (and hence greater vorticities) are accessible. As a result of the formation of smaller scales in the t5 experiment (in comparison to the s5 case), we expect an additional increase in the efficiency of wave energy dissipation compared to the uniform field case in non-ideal simulations.

\begin{figure}[h]
   \includegraphics[width=0.45\textwidth]{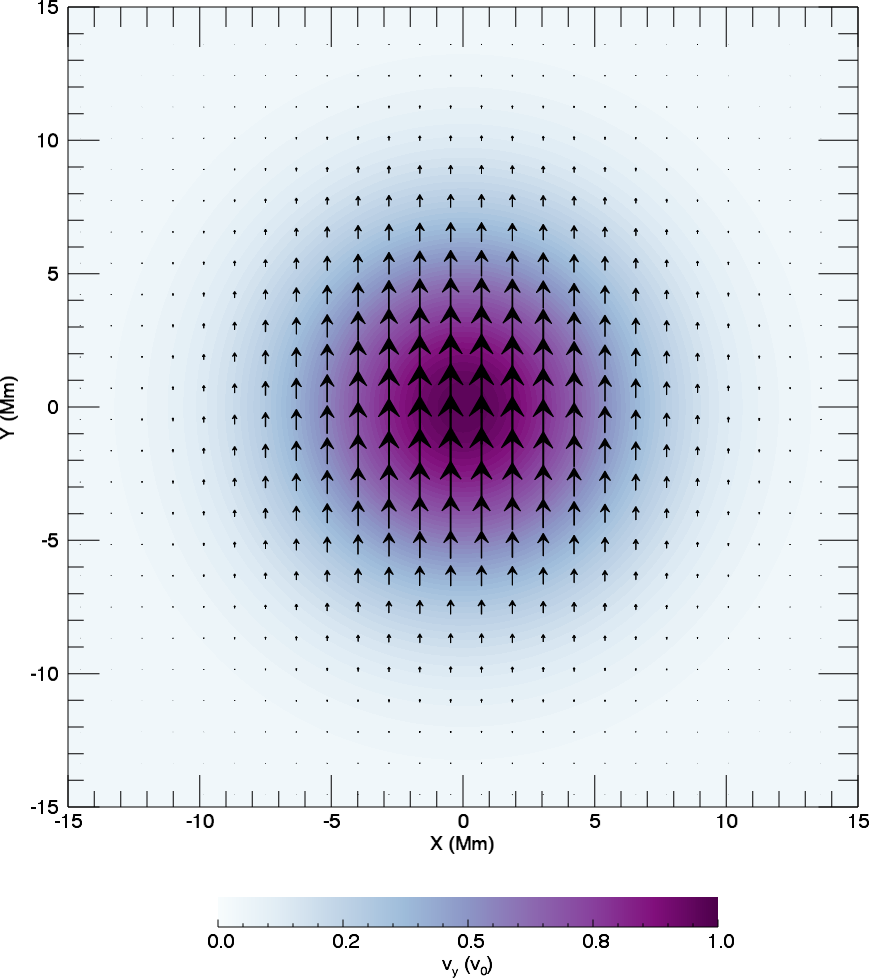}
   \caption{Non-uniform driver profile imposed at the lower $z$ boundary.}
   \label{spat_driver} 
\end{figure}

\subsection{Non-uniform wave driver}\label{non-unidriver}
In this section, we consider the effects of imposing a transverse wave driver that is not spatially uniform across the lower boundary of the domain. As waves propagate along magnetic structures, the complexity of the field ensures that small gradients at the foot points of field lines can map to large gradients within the numerical domain. It is reasonable to expect that two field lines which are well-separated within the lower solar atmosphere may be excited by different wave drivers. If these field lines converge towards each other at higher altitudes, then as the wave packets propagate into the corona, the transverse gradients will be enhanced. 

We consider a modified driver of the form 
\begin{equation}v_y = \label{wd_eqn2}\begin{cases}
v_0\sin\left(\frac{2\pi t}{\tau}\right)e^{-\left(^{r}/_{l}\right)^2} \hspace{1.5cm} &\text{if } t \le \tau,\\
0  \hspace{1.5cm} &\text{if } t > \tau,  
\end{cases}
\end{equation}
where $r^2 = x^2 + y^2$. We set $l=6$ Mm to ensure that the imposed velocity is approximately $0$ close to the $x$ and $y$ boundaries of the domain. The form of the imposed wave driver is shown in Fig. \ref{spat_driver} and the total Poynting flux through the lower boundary of the domain is shown in Fig. \ref{Poynting_injection} (dashed red line). It is imposed on two initial conditions; the s5 field and, for comparison, the straight field case.

\begin{figure}[h]
   \includegraphics[width=0.45\textwidth]{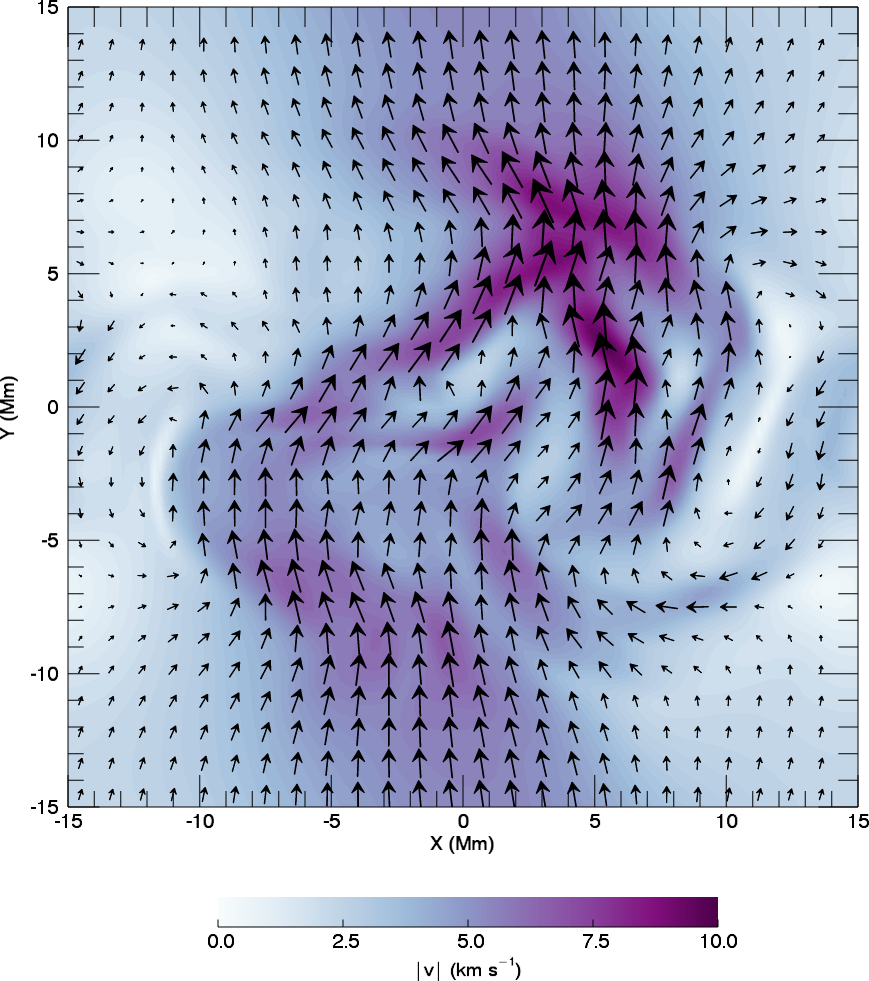}
   \caption{Transverse velocity profile in a cross-section of the wave packet at $t=0.8\,T_e$ for the non-uniform driver.}
   \label{spat_wave} 
\end{figure}

In Fig. \ref{spat_wave}, we show the transverse velocity profile in a horiztontal cross-section through the wave front at $t=0.8\,T_e$. We show the magnitude of the horizontal velocity perturbation (contour plot) and the direction of these flows (overplotted vectors). We note that although the wave driver has the same maximum amplitude as previously, the flows are smaller in this case. This is because the perturbed plasma has to displace surrounding fluid that is not moved by the imposed velocity driver. This transfers wave energy from the centre of the domain to the surrounding plasma. Additionally, there is some evidence of a return flow being generated at large $\lvert x \rvert$ as the non-driven plasma moves into the space vacated by the perturbed fluid. 

The small scales observed in Fig. \ref{spat_wave} are once again indicative of the phase mixing that progresses as the wave propagates along magnetic field lines. However, since the wave energy injected by the driver is smaller in this case (compare the dashed and solid red lines in Fig. \ref{Poynting_injection}), the total vorticity is also reduced. Therefore, in order to compare the growth in small scales due to phase mixing for the two wave drivers, we normalise the vorticity by the maximum value obtained in the corresponding straight field simulations.

\begin{figure}[h]
   \includegraphics[width=0.5\textwidth]{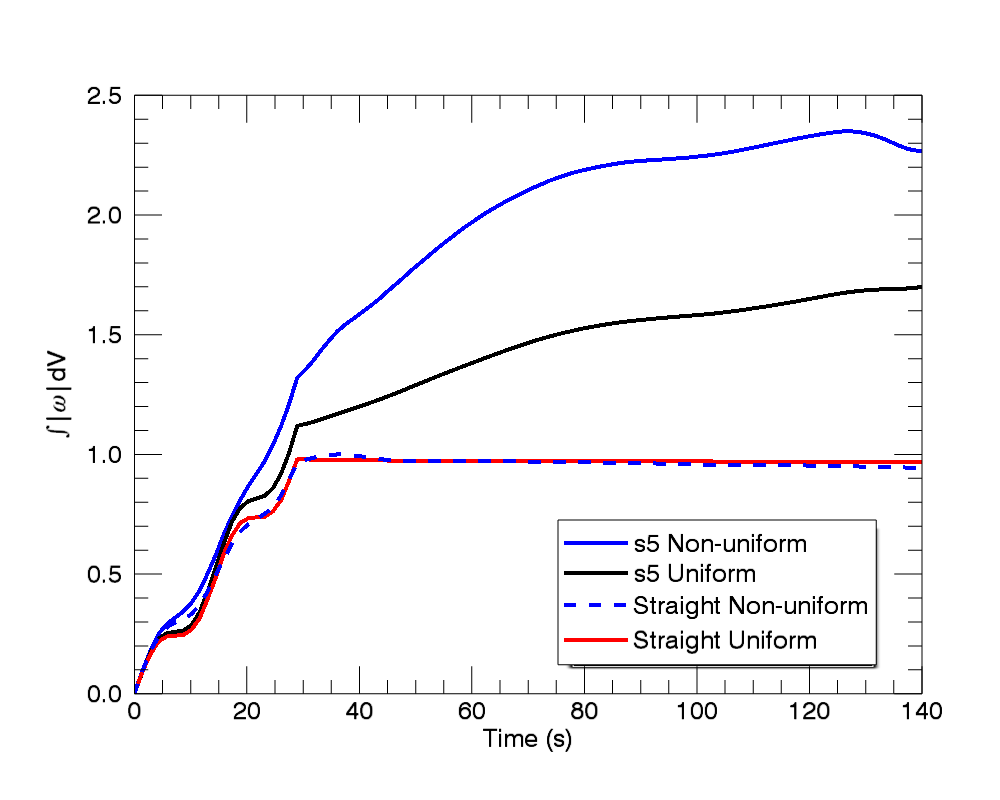}
   \caption{Volume integrated $\lvert \omega \rvert$ for the two forms of the wave driver in the s5 and straight field simulations. For the uniform driver cases, we have normalised by the maximum of the red curve. For the non-uniform driver cases, we have normalised by the maximum of the dashed blue curve.}
   \label{spat_vort_comparison} 
\end{figure}

In Fig. \ref{spat_vort_comparison}, we show the magnitude of the vorticity, integrated over the computational domain for the two forms of the driver. We show the uniform driver acting on the straight field and s5 initial conditions (red and black lines, respectively). For both of these curves, we normalise by the maximum of the red line. Additionally, we show the non-uniform driver on these two initial conditions (dashed and solid blue lines, respectively). For these two curves, we have normalised by the maximum of the dashed blue line.

In Fig. \ref{spat_vort_comparison}, we see that once the reduction in the injected vorticity (for the non-uniform driver) is accounted for, the integral of $\lvert \omega \rvert$ increases by a greater amount in the non-uniform cases. This is largely due to the phenomenon described above by which complex magnetic fields can ensure small gradients in the wave driver can map to large gradients in the numerical domain. As such, in a non-ideal regime, wave energy dissipation will be significantly more efficient for a wave driver that has (possibly small) spatial gradients. 

%%%%%%%%%%%%%%%%%%%%%%%%%%%%%%%%%%%%%%%%%%%%%%%%%%%%%%%%%%%%%%%%%%%%%%%
%Discussion and Conclusions
%%%%%%%%%%%%%%%%%%%%%%%%%%%%%%%%%%%%%%%%%%%%%%%%%%%%%%%%%%%%%%%%%%%%%%%
\section{Discussion and conclusions}\label{Sect_Discussion}
In this article we have presented the results of MHD simulations of transverse wave packets propagating through complex magnetic field configurations. A single sinusoidal pulse was driven on the lower boundary of the domain for initial equilibria containing varying field complexity. 

The presence of an approximately helical component in the magnetic field induces a localised change in the polarisation of the wave front, which is sensitive to the complexity of the initial currents. Additionally, the inhomogeneous Alfv\'en speed profile associated with the background plasma and magnetic field generates a complex phase mixing pattern. Enhancing the field complexity increases the average rate of phase mixing and the formation of small scales throughout the wave front.  

As a result of these effects, large spatial gradients are induced in both the perturbed velocity and magnetic fields. However, the pre-existing field complexity ensures it is relatively difficult to identify the formation of additional small scales in the magnetic field as the wave propagates through the numerical domain. On the other hand, the background flows are very small (in comparison to the driver amplitude) and thus we have focussed our analysis on the evolution of the velocity field. In particular, we highlight the evolution of the vorticity as a useful measure for tracking the generation of small scales.

The single pulse examined in the current study contains very little energy and thus, even in non-ideal regimes, does not lead to significant plasma heating. The Poynting flux could certainly be enhanced by adopting a larger amplitude driver or by increasing the background field strength. Alternatively, a continuous driver will inject substantially more energy into the domain and the potential excitation of field line resonances can further increase the energy input. Since the field exhibits a wide range of natural field line frequencies, it is relatively easy to excite resonances regardless of the driving frequency. The additional energy input and the potential for even more complex dynamics ensures wave heating is more significant in a continuously driven case. A more thorough investigation of such simulations will be considered in subsequent work.

We have discussed the enhanced efficiency of kinetic energy dissipation in non-ideal regimes within the complex field structures. In a relatively low Reynolds number plasma, the increase in heating is limited in comparison to the dissipation associated with an Alfv\'en wave propagating through a uniform field. However, at higher Reynolds numbers, the enhancement is greater (compared to an analagous uniform field case) due to the increased effects of phase mixing for reduced viscosities. Additionally, the implementation of a non-uniform driver generates further small scales due to the convergence of initially distant magnetic field lines within the corona. As such, in the presence of continuous wave driving, this model may allow significant heat to be deposited within the solar atmosphere. At present, this hypothesis remains untested and will be the focus of future research.

Many classical phase mixing models rely on a prescribed density profile which are not self-consistently generated by the heating process. Indeed, \citet{Cargill2016} argued that even with transport coefficients augmented by many orders of magnitude above expected coronal values, the heating profile established by phase mixing will not be able to sustain the dense loops observed within the solar atmosphere. Since, in this model, the spatial distribution of energy dissipation is not confined to regions of strong density gradient, it seems that the arguments presented by \citet{Cargill2016} do not preclude phase mixing as a coronal heating mechanism in this regime. It is important to highlight that, in these simulations, wave energy is dissipated across the entire cross-section of the braided magnetic structure and not simply in narrow layers within the flux tubes.

Since the transverse wave propagates through a highly inhomogeneous domain, it cannot be identied as a pure Alfv\'en wave. Indeed, in this regime the wave is weakly compressible and, as such, given sufficient telescope sensitivity, may be directly observable in the solar corona. Subsequent work will seek to identify the necessary wave amplitude for such plasma compression to be detected with current observational capabilities. 

In any case, the wave front becomes highly deformed and transfers energy to smaller scales which will be beyond the spatial resolving power of observing instruments. As such, it remains unclear whether the localised phase mixing pattern and modification of the wave polarisation have implications for estimating the coronal wave energy budget. In particular, it may be difficult to accurately estimate the energy associated with observations of similar waves in the solar atmosphere.

The compressibility of the propagating wave front is not homogeneous and is sensitive to the nature of the background field. For example, it is associated with greater density variation in regions of greater field complexity. Therefore, it is plausible that observations of the weak plasma compression as a similar wave propagates through coronal field will indicate the complexity of the background Alfv\'en speed and/or the nature of magnetic field lines. As such synthetic observables derived from the models presented within this article are expected to contain information about the initial conditions. It remains to be seen whether any seismological techniques can be employed on the wave dynamics to infer properties of the background medium. If so, observations of propagating transverse coronal waves may be used to infer the complexity of the coronal magnetic field. \\

{\emph{Acknowledgements.}} The authors would like to thank the referee for their helpful comments. The research leading to these results has received
funding from the UK Science and Technology Facilities Council (consolidated
grants ST/N000609/1 and ST/S000402/1) and the European Union Horizon 2020 research
and innovation programme (grant agreement No. 647214). IDM acknowledges support from the Research Council of Norway through its Centres of Excellence scheme, project number 262622 and JR acknowledges the support of the Carnegie Trust for the Universities of Scotland.

\bibliographystyle{aa}        % suitable bib style file 
\bibliography{WBF.bib}           % include your own .bib file

\end{document}